\documentclass[aps,pra,amssymb,twocolumn,showpacs,floatfix]{revtex4}
\usepackage{graphicx,amssymb,amsmath,bbm,amsfonts}
\usepackage{amsmath}
\usepackage{amsfonts}
\usepackage{amssymb}
\usepackage{graphicx}
\begin{document}
\title{Optimal detection of losses by thermal probes}
\author{Carmen Invernizzi}
\email{Carmen.Invernizzi@unimi.it}
\affiliation{Dipartimento di Fisica dell'Universit\`a degli Studi di Milano, I-20133
Milano, Italy}
\author{Matteo G. A. Paris}
\email{matteo.paris@fisica.unimi.it}
\affiliation{Dipartimento di Fisica dell'Universit\`a degli Studi di Milano, I-20133
Milano, Italy}
\affiliation{CNISM, Udr Milano, I-20133 Milano, Italy}
\author{Stefano Pirandola}
\email{pirs@cs.york.ac.uk}
\affiliation{Department of Computer Science, University of York, York YO10 5GH, United Kingdom}

\begin{abstract}
We consider the discrimination of lossy bosonic channels and focus to the case
when one of the values for the loss parameter is zero, i.e., we address the
detection of a possible loss against the alternative hypothesis of an ideal
lossless channel. This discrimination is performed by inputting one-mode or
two-mode squeezed thermal states with fixed total energy. By optimizing over
this class of states, we find that the optimal inputs are pure, thus
corresponding to single- and two-mode squeezed vacuum states. In particular,
we show that for any value of the damping rate smaller than a critical value
there is a threshold on the energy that makes the two-mode squeezed vacuum
state more convenient than the corresponding single-mode state, whereas for
damping larger than this critical value two-mode squeezed vacua are always
better. We then consider the discrimination in realistic conditions, where it
is unlikely to have pure squeezing. Thus by fixing both input energy and
squeezing, we show that two-mode squeezed thermal states are always better
than their single-mode counterpart when all the thermal photons are directed
into the dissipative channel. Besides, this result also holds approximately
for unbalanced distribution of the thermal photons. Finally, we also
investigate the role of correlations in the improvement of detection. For
fixed input squeezing (single-mode or two-mode), we find that the reduction of
the quantum Chernoff bound is a monotone function of the two-mode entanglement
as well as the quantum mutual information and the quantum discord. We thus
verify that employing squeezing in the form of correlations (quantum or
classical) is always a resource for loss detection whenever squeezed thermal
states are taken as input.

\end{abstract}

\date{\today}
\maketitle

\section{Introduction}

One of the main obstacles to the development of quantum technologies is the
decoherence associated to losses and absorption processes occurring during the
propagation of a quantum signal. The description of the dynamics of systems
subject to noisy environments \cite{Breuer}, as well the detection,
quantification and estimation of losses and, more generally, the
characterization of lossy channels at the quantum level, received much
attention in the recent years \cite{Breuer, DecoRev, Dau06, Zurek}. An
efficient characterization of decoherence is relevant for quantum repeaters
\cite{Brask}, quantum memories \cite{Jensen}, cavity QED systems
\cite{Haroche08}, superconducting quantum circuits \cite{Wang}, quantum
teleportation \cite{Qtele}, quantum cryptography \cite{QKD1,QKD2} and
secret-key capacities \cite{Devetak,PirSKcapacity}.

In this paper we address the discrimination of lossy channels, i.e., we
consider a situation where the loss (damping) rate of a channel may assume
only two possible values and we want to discriminate between them by probing
the channel with a given class of states. In particular, we address the
discrimination of lossy channels for bosonic systems using squeezed thermal
states as probing states, and focus attention to the case when one of the
values for the loss parameter is zero, i.e., we address the detection of a
possible loss against the alternative hypothesis of an ideal lossless channel.
Such kind of discrimination is crucial since a recent analysis \cite{Mon11}
has revealed the importance of assessing the deviation from ideal conditions,
i.e., the identity channel, in implementing large-scale quantum communication.

Despite the discrimination of lossy channels has been already considered in
the literature, the approach of this paper is novel. Previous works have in
fact considered this kind of discrimination by constraining the energy
irradiated over the unknown lossy channel but not the total energy employed in
preparing the input states. For instance, in the quantum reading of digital
memories~\cite{QreadingPRL}, the discrimination of lossy channels has been
analyzed by fixing the mean total number of photons irradiated over the
channel, independently on the number of probing modes (this approach has been
also followed by the recent Ref.~\cite{NairLAST}). Previously, in the quantum
illumination of targets~\cite{Qillumination,QilluminationOTHERS}, the channel
discrimination was performed by fixing the mean number of photons in each of
the modes probing the channel. In both these models there was no restriction
for the energy involved in the use of ancillary modes. Our approach considers
the discrimination of lossy channels by constraining the total energy of the
input state, thus including both the probing mode (irradiated over the unknown
lossy channel) and a possible ancillary mode (bypassing the lossy channel and
detected by the output measurement). Thus, while previous models were more
focussed on restricting the energy irradiated over the channel, we address the
problem from the point of view of the input source, i.e., considering the
global effort in preparing this source. By fixing the input energy, we then
optimize over an important class of Gaussian states, i.e., single- and
two-mode squeezed thermal states. The choice of these states relies in their
experimental accessibility, being routinely generated in today's quantum
optics laboratories where they can be reliably controlled \cite{PRLtwo-mode}.
Furthermore, because of the squeezing, they represent important examples of
non-classical states, i.e., states with non-positive P representation
\cite{Glauber}. This is another feature which diversifies our study from
previous works~\cite{QreadingPRL,Qillumination}, where the main goal was the
comparison between non-classical states and classical states (i.e., with
positive P representation). Note also that our problem, involving a discrete
channel discrimination, is completely different from problems of parameter
estimation, where one has to infer a parameter taking continuous values. The
estimation of the damping constant of a bosonic channel has been recently
addressed by Refs.~\cite{Freyberger, Mon07, AdessoR,Mon10}.

It is clear that, for a given input state, any problem of channel
discrimination collapses into a problem of state discrimination
\cite{Helstrom, Chefles, QSE, QSE2, Kargin}, where we have to compute the
minimum error probability in identifying one of two possible output states. By
assuming $M$ identical copies of the input state and a memoryless quantum
channel, we have $M$ output states which are exact replicas of two
equiprobable states. In this case, the minimum error probability is
well-approximated by the quantum Chernoff bound (QCB) \cite{PRAQCB,
PRLQCB,Nuss,Aud,Pirandola}. Now the crucial step is to vary the input state,
trying to optimize the value of the output QCB. In the case of two lossy
bosonic channels, this kind of optimization must be constrained, meaning that
we have to fix some crucial parameters of the input states, in particular,
their energy. As we have already mentioned before, in our investigation we
optimize the QCB on the class of single- and two-mode squeezed thermal states
by fixing their total energy. Single-mode thermal states are sent through the
lossy channel, while two-mode thermal states probe the channel with the one of
the modes (probing mode) while bypassing the channel and assisting the output
measurement with the other mode (reference mode). In this scenario, we find
that the pure version of these states, i.e., single- and two-mode squeezed
vacuum states, are optimal for detecting losses, i.e., discriminating a lossy
from a lossless channel. Furthermore, we are able to show that, for any value
of the damping rate smaller than a certain critical value, there is a
threshold on the energy that makes the two-mode squeezed vacuum state more
convenient than the corresponding single-mode state. More interestingly, for
damping rates larger than the critical value, the two-mode squeezed vacuum
state performs always better than the single-mode squeezed vacuum state with
exactly the same energy.

In order to stay close to schemes which are feasible with current technology,
we also analyze the effect of the mixedness in the probe states. In this case,
we study the channel discrimination by fixing not only the total energy of the
input state but also its total amount of squeezing. Then, we are able to show
that two-mode squeezed thermal states are always better than their single-mode
counterpart when all the thermal photons are directed into the dissipative
channel. We have numerically verified that this result also holds,
approximately, for unbalanced distribution of the thermal photons. Finally, we
also investigate the role of correlations in the improvement of loss
detection. In order to quantify correlations, besides entanglement and mutual
information, we exploit the recent results on the quantum
discord~\cite{Disc1,Disc2,Ferraro10,GQDSCa,GQDSCb,Vedral}, which has been
defined with the aim of capturing quantum correlations in mixed separable
states that are not quantified by entanglement. Thus, at fixed input
squeezing, we study the reduction of the QCB as a function of various
correlation quantifiers, i.e., quantum mutual information, entanglement and
quantum discord. This analysis allows us to conclude that employing squeezing
in the form of correlations, either quantum or classical, is beneficial for
the task of loss detection whenever squeezed thermal states are considered as
input probes.

The paper is structured as follows. In Sec.~\ref{QHT} we review the
discrimination of quantum states together with the general definition of the
QCB. In Sec.~\ref{QCBinGS} we give the basic notation for Gaussian states and
in Sec.~\ref{s:qcbG} we review the formula of the QCB for Gaussian states.
Then, in Sec.~\ref{QCBloss} we discuss the discrimination of lossy channels by
considering single-mode and two-mode squeezed thermal states. In particular,
we provide the details on how to compute the QCB for distinguishing an ideal
lossless channel from a lossy channel. Section \ref{results} reports the main
results regarding the single- and two-mode states as a function of the total
energy and squeezing. Finally, in Section \ref{qcbcorr} we analyze the role of
correlations in enhancing the discrimination. Section \ref{conclusions} closes
the paper with some concluding remarks.
%%%%%%%

\section{Quantum State Discrimination and Quantum Chernoff Bound}

\label{QHT} As we have mentioned before, the problem of quantum channel
discrimination collapses to the problem of quantum state discrimination when
we fix the input state. As a result, mathematical tools such as the quantum
fidelity and the QCB are fundamental in our analysis. Despite it has been
introduced only very recently, the QCB has been already a crucial tool in
several areas of quantum information: it has been exploited as a
distinguishability measure between qubits and single-mode Gaussian states
\cite{PRAQCB, PRLQCB}, to evaluate the degree of nonclassicality for
single-mode Gaussian states \cite{Boca} or the polarization of a two-mode
state \cite{Ghiu}. It has also been applied in the theory of quantum phase
transitions to distinguish between different phases of the $XY$ model at
finite temperature \cite{AbastoQCM}, and to the discrimination of two ground
states or two thermal states in the quantum Ising model \cite{JMOQCB}. For
continuous variables systems the quantum discrimination of Gaussian states is
a central point in view of their experimental accessibility and their
relatively simple mathematical description \cite{GSinQI, QITCV}. In fact, in
the case of Gaussian states, the QCB can be computed from their first and the
second statistical moments. A first formula, valid for single-mode Gaussian
states, was derived in \cite{PRAQCB}. Later, Ref.~\cite{Pirandola} provided a
general closed formula for multimode Gaussian states, relating the QCB bound
to their symplectic spectra. Furthermore, from these spectra one can derive
larger upper bounds which are easier to compute than the QCB \cite{Pirandola}.

In this section, we start by establishing notation and reviewing the problem
of quantum state discrimination, together with the general definition of QCB.
Then, from the next section, we will specialize our attention to the case of
Gaussian states and we will review the formula for Gaussian states in
Sec.~\ref{s:qcbG}.

In its simplest formulation, the problem of quantum state discrimination
consists in distinguishing between two possible states, $\rho_{A}$ and
$\rho_{B}$, which are equiprobable for a quantum system. We suppose that $M$
identical copies of the quantum system are available. Then, we have the
following equiprobable hypotheses, $H_{A}$ and $H_{B}$, about the global
state
\begin{align*}
H_{A} &  :\rho_{A}^{M}=\underbrace{\rho_{A}\otimes\ldots\otimes\rho_{A}}_{M}\\
H_{B} &  :\rho_{B}^{M}=\underbrace{\rho_{B}\otimes\ldots\otimes\rho_{B}}_{M}.
\end{align*}
In order to discriminate between these two hypotheses, one can measure the
global system by using a two-outcome positive operator valued measure (POVM)
$\{E_{A},E_{B}\}$, with $E_{A}+E_{B}={\mathbb{I}}$ and $E_{A},E_{B}\geq0$.
After observing the outcome $j=A$ or $B$, the observer infers that the state
of the system was $\rho_{j}^{M}$. The error probability of inferring the state
$\rho_{j}^{M}$ when the actual state is $\rho_{k}^{M}$ is thus given by the
Born rule $P_{jk}=\hbox{Tr}\left[  \rho_{k}^{M}E_{j}\right]  $. As a result,
the optimal POVM for this discrimination problem is the one minimizing the
overall probability of misidentification, i.e., $P_{e}=\frac{1}{2}%
(P_{BA}+P_{AB})$. Since $E_{A}={\mathbb{I}}-E_{B}$, we have
\begin{align}
P_{e} &  =\frac{1}{2}\hbox{Tr}[\rho_{A}^{M}E_{B}]+\frac{1}{2}\hbox{Tr}[\rho
_{B}^{M}E_{A}]\nonumber\\
&  =\frac{1}{2}\left(  1-\hbox{Tr}\left[  E_{B}\Lambda\right]  \right)  ~,
\end{align}
where
\[
\Lambda=\rho_{B}^{M}-\rho_{A}^{M}\,,
\]
is known as the Helstrom matrix~\cite{Helstrom}. Now, the error probability
$P_{e}$ has to be minimized over $E_{B}$. Since $\hbox{Tr}[\Lambda]=0$, the
Helstrom matrix has both positive and negative eigenvalues and the minimum
$P_{e}$ is attained if $E_{B}$ is chosen as the projector over $\Lambda_{+}$,
i.e., the positive subspace of $\Lambda$. Assuming this optimal operator we
have $\hbox{Tr}[E_{B}\Lambda]=\mathop{\text{Tr}}\nolimits[\Lambda_{+}%
]=\frac{1}{2}\mathop{\text{Tr}}\nolimits|\Lambda|$ with $|\Lambda
|=\sqrt{\Lambda^{\dagger}\Lambda}$. Thus the minimal error probability is
given by
\[
P_{e}=\frac{1}{2}\left[  1-T(\rho_{A}^{M},\rho_{B}^{M})\right]  ~,
\]
where
\[
T(\rho,\sigma)=\frac{1}{2}\mathop{\text{Tr}}\nolimits|\rho-\sigma|
\]
is the so-called trace distance. The computation of the trace distance may be
rather difficult. For this reason, one can resort to the QCB that gives an
upper bound to the probability of error $P_{e}$ \cite{PRAQCB,
PRLQCB,Nuss,Aud,Pirandola}%
\begin{equation}
P_{e}\leq\frac{Q^{M}}{2}~,\label{bound:Pe}%
\end{equation}
where
\begin{equation}
Q=\inf_{0\leq s\leq1}\mathop{\text{Tr}}\nolimits\left[  \rho_{A}^{s}\rho
_{B}^{1-s}\right]  ~.\label{QCBgen}%
\end{equation}
The bound of Eq.~(\ref{bound:Pe}) is attainable asymptotically in the limit
$M\rightarrow\infty$ as follows from the results in \cite{PRLQCB,Nuss}. One
may think that the trace distance has a more natural operational meaning than
the QCB. In spite of this, it does not adapt to the case of many copies;
indeed, one can find states $\rho,\sigma,\rho^{\prime},\sigma^{\prime}$ such
that
\[
T(\rho,\sigma)<T(\rho^{\prime},\sigma^{\prime})\quad\hbox{but}\quad
T(\rho^{\prime}{}^{M},\sigma^{\prime}{}^{M})<T(\rho^{M},\sigma^{M})\,.
\]
By contrast, the QCB does resolve this problem since
\[
Q(\rho,\sigma)<Q(\rho^{\prime},\sigma^{\prime})\;\Longrightarrow\;Q(\rho
^{M},\sigma^{M})<Q(\rho^{\prime M},\sigma^{\prime M})\,.
\]
Because of this property, the minimization of the QCB over single-copy states
($\rho$ and $\sigma$) implies the minimization over multi-copy states
($\rho^{M}$ and $\sigma^{M}$). This is true as long as the minimization is
unconstrained or if the constraints regard single-copy observables (e.g., the
mean energy per copy).

Finally, note that there is a close relation between the QCB and the Uhlmann
fidelity%
\[
F(\rho_{A},\rho_{B})=\mathop{\text{Tr}}\nolimits\left(  \sqrt{\sqrt{\rho_{A}%
}\rho_{B}\sqrt{\rho_{A}}}\right)  ^{2}%
\]
which is one of the most popular measures of distinguishability for quantum
states. In fact, for the single-copy state discrimination ($M=1$)\ we
have~\cite{PRAQCB,Fuchs,Nielsen,Pirandola}%
\begin{equation}
\frac{1-\sqrt{1-F(\rho_{A},\rho_{B})}}{2}\leq P_{e}\leq\frac{Q}{2}\leq
\frac{\sqrt{F(\rho_{A},\rho_{B})}}{2}.\label{1copyINEQ}%
\end{equation}
More generally, by exploiting the multiplicativity of the fidelity under
tensor products of density operators, i.e.,%
\[
F(\rho_{A}\otimes\sigma_{A},\rho_{B}\otimes\sigma_{B})=F(\rho_{A},\rho
_{B})F(\sigma_{A},\sigma_{B})~,
\]
we can write
\[
F(\rho_{A}^{M},\rho_{B}^{M})=F(\rho_{A},\rho_{B})^{M}=F^{M}~.
\]
This leads to the general multi-copy version of Eq.~(\ref{1copyINEQ}) which is
given by~\cite{QreadingSUPP}%
\begin{equation}
\frac{1-\sqrt{1-F^{M}}}{2}\leq P_{e}\leq\frac{Q^{M}}{2}\leq\frac{F^{M/2}}{2}~.
\end{equation}
From the previous inequalities, it is clear that the QCB\ gives a tighter
bound than the quantum fidelity. However, if one of the two states is pure,
then the QCB just equals the fidelity, i.e., we have
\[
Q(\rho_{A},\rho_{B})=F(\rho_{A},\rho_{B})=\mathop{\text{Tr}}\nolimits[\rho
_{A}\,\rho_{B}]~.
\]

\section{Gaussian states}

\label{QCBinGS} In this section we give the basic notions on bosonic systems
and Gaussian states, ending with the definition of squeezed thermal states.

An $n$-mode bosonic system is described by a tensor-product Hilbert space
$\mathcal{H}^{\otimes n}$ and a vector of canonical operators $\boldsymbol{R}%
=(q_{1},p_{1},\ldots,q_{n},p_{n})^{T}$ satisfying the commutation relations
\[
\lbrack R_{l},R_{m}]=i\Omega_{lm}~,
\]
where $l,m=1,\cdots,2n$ and $\Omega_{lm}$ are the elements of the symplectic
form
\begin{equation}
\boldsymbol{\Omega}=\bigoplus_{k=1}^{n}\left(
\begin{array}
[c]{cc}%
0 & 1\\
-1 & 0
\end{array}
\right)  ~. \label{syForm}%
\end{equation}
Alternatively, we can use the mode operators $a_{k}$ which are given by the
cartesian decomposition of the canonical operators, i.e.,%
\[
q_{k}=\frac{1}{\sqrt{2}}(a_{k}+a_{k}^{\dagger})~,~p_{k}=\frac{1}{i\sqrt{2}%
}(a_{k}-a_{k}^{\dagger})~.
\]
These operators satisfy the commutation relations $[a_{k},a_{k^{\prime}%
}^{\dagger}]=\delta_{kk^{\prime}}$ with $k,k^{\prime}=1,\cdots,n$.

An arbitrary quantum state $\rho$ of the system is equivalently described by
the characteristic function
\[
\chi\lbrack\rho](\boldsymbol{\lambda})=\mathop{\text{Tr}}\nolimits[\rho
D(\boldsymbol{\lambda})]
\]
where $D(\boldsymbol{\lambda})=\otimes_{k=1}^{n}D_{k}(\lambda_{k})$ is the
$n$-mode displacement operator, with $\boldsymbol{\lambda}=(\lambda_{1}%
,\ldots,\lambda_{n})^{T}$, $\lambda_{k}\in\mathbb{C}$, and $D_{k}(\lambda
_{k})=\exp\{\lambda_{k}a_{k}^{\dagger}-\lambda_{k}^{\ast}a_{k}\}$ is the
single mode displacement operator. A state $\rho$ is called Gaussian if the
corresponding characteristic function is Gaussian
\begin{equation}
\chi\lbrack\rho](\boldsymbol{\Lambda})=\exp\left\{  -\frac{1}{2}%
\boldsymbol{\Lambda^{T}\sigma\Lambda}+\boldsymbol{X^{T}\Omega\Lambda}\right\}
\end{equation}
where $\boldsymbol{\Lambda}$ is the real vector%
\[
\boldsymbol{\Lambda}=(\text{\textrm{Re}}\lambda_{1},\text{\textrm{{Im}}%
}\lambda_{1}\mathrm{,\ldots,{Re}}\lambda_{n}\mathrm{,{Im}}\lambda_{n})^{T}~.
\]
In this case, the state is described by its first two statistical moments,
i.e., the vector of mean values $\boldsymbol{X}$ and the covariance matrix
(CM) $\boldsymbol{\sigma}$, whose elements are defined as
\begin{align}
X_{l}=  &  \langle R_{l}\rangle\nonumber\\
\sigma_{lm}=  &  \frac{1}{2}\langle\{R_{l},R_{m}\}\rangle-\langle R_{l}%
\rangle\langle R_{m}\rangle
\end{align}
where $\{A,B\}=AB+BA$ denotes the anti-commutator, and $\langle O\rangle
=\mathop{\text{Tr}}\nolimits[\rho O]$ is the mean value of the operator $O$.

In the remainder of this section, we consider only zero-mean Gaussian states,
i.e., Gaussian states with $\boldsymbol{X}=0$, which are therefore fully
specified by their CM. The properties of these states may be expressed in very
simple terms by introducing the symplectic transformations. A matrix
$\boldsymbol{S}$ is called symplectic when preserves the symplectic form of
Eq.~(\ref{syForm}), i.e.,
\[
\boldsymbol{S\Omega S^{T}}=\boldsymbol{\Omega}.
\]
Then, according to the Williamson's theorem, for every CM $\boldsymbol{\sigma
}$, there exists a symplectic matrix $\boldsymbol{S}$ such that
\begin{equation}
\boldsymbol{\sigma}=\boldsymbol{SWS^{T}}\label{syDECOMP}%
\end{equation}
where
\[
\boldsymbol{W}=\bigoplus_{k=1}^{n}d_{k}\left(
\begin{array}
[c]{cc}%
1 & 0\\
0 & 1
\end{array}
\right)  ~,
\]
and the $d_{k}$'s are called the symplectic eigenvalues of $\boldsymbol{\sigma
}$. The physical statement implied by the decomposition of Eq.~(\ref{syDECOMP}%
) is that every zero-mean Gaussian state $\rho$ can be obtained from a thermal
state by performing the unitary transformation $U_{\boldsymbol{S}}$ associated
with the symplectic matrix $\boldsymbol{S}$, i.e., we have
\[
\rho=U_{\boldsymbol{S}}~\boldsymbol{\nu~}U_{\boldsymbol{S}}^{\dag}%
\]
where $\boldsymbol{\nu}=\nu_{1}\otimes\ldots\otimes\nu_{n}$ is a tensor
product of single-mode thermal states
\[
\nu_{k}=\frac{1}{\bar{n}_{k}+1}\sum_{m}\left(  \frac{\bar{n}_{k}}{\bar{n}%
_{k}+1}\right)  ^{m}|m\rangle_{k}\langle m|
\]
with average number of photons given by $\bar{n}_{k}=d_{k}-1/2$. For a
single-mode system the most general zero-mean Gaussian state may be written
as
\[
\rho=S(\zeta)\nu S^{\dagger}(\zeta)
\]
where $S(\zeta)=\exp\{\frac{1}{2}({\zeta a^{\dagger}}^{2}-\zeta^{\ast}{a}%
^{2})\}$ is the single-mode squeezing operator and $\zeta=re^{i\phi}%
\in\mathbb{C}$. The corresponding covariance matrix is given by%
\begin{equation}
\boldsymbol{\sigma}=\left(
\begin{array}
[c]{cc}%
a & c\\
c & b
\end{array}
\right)  \label{CM_1}%
\end{equation}
where
\begin{align}
a &  =(\bar{n}+\frac{1}{2})\left[  \cosh(2r)-\sinh(2r)\cos\phi\right]
\nonumber\\
b &  =(\bar{n}+\frac{1}{2})\left[  \cosh(2r)+\sinh(2r)\cos\phi\right]
\nonumber\\
c &  =(\bar{n}+\frac{1}{2})\sinh(2r)\sin\phi~.\label{Expressions}%
\end{align}
In particular, we can consider the case of a real squeezing parameter, e.g.,
by fixing $\zeta=-r$~\cite{Segno}. In this case, the previous expressions of
Eq.~(\ref{Expressions}) simplify into the following%
\begin{align}
a &  =\frac{1}{2}(2\bar{n}+1)\exp(2r)\nonumber\\
b &  =\frac{1}{2}(2\bar{n}+1)\exp(-2r)\nonumber\\
c &  =0~.\label{ExpressionsSIMPLE}%
\end{align}
This state defines the single-mode squeezed thermal state. It depends on two
real parameters only, i.e., we have
\[
\rho=S(r)\nu S^{\dagger}(r)=\rho(r,\bar{n})~.
\]
In particular, for $\bar{n}=0$ the state is pure and corresponds to a
single-mode squeezed vacuum state $\rho(r,0)=S(r)\left\vert 0\right\rangle
\left\langle 0\right\vert S^{\dagger}(r)$.

Now let us consider two-mode (zero-mean) Gaussian states. They are completely
characterized by their $4\times4$ CM%
\begin{equation}
\boldsymbol{\sigma}=\left(
\begin{array}
[c]{cc}%
\mathbf{A} & \mathbf{C}\\
\mathbf{C}^{T} & \mathbf{B}%
\end{array}
\right)  \label{CM_2modes}%
\end{equation}
where $\mathbf{A}$, $\mathbf{B}$ and $\mathbf{C}$\ are $2\times2$ blocks. It
is useful to introduce the symplectic invariants
\begin{align}
I_{1}  &  =\det\mathbf{A}~,~I_{2}=\det\mathbf{B}\,,\nonumber\\
I_{3}  &  =\det\mathbf{C}~,~I_{4}=\det\boldsymbol{\sigma}~.
\label{syINVARIANTS}%
\end{align}
By means of these invariants, we can simply write the two symplectic
eigenvalues as%
\[
d_{\pm}=\sqrt{\frac{\Delta\pm\sqrt{\Delta^{2}-4I_{4}}}{2}}~,
\]
where $\Delta=I_{1}+I_{2}+2I_{3}$~\cite{DiagSympl,PirCMs}. By means of local
symplectic operations, the CM of Eq.~(\ref{CM_2modes}) can be recast in the
standard form, where the three blocks $\mathbf{A}$ and $\mathbf{B}$ are
proportional to the identity and $\mathbf{C}$\ is diagonal~\cite{DiagSympl}.
In the particular case of a two-mode squeezed thermal state, we can write%
\begin{equation}
\boldsymbol{\sigma}=\frac{1}{2}\left(
\begin{array}
[c]{cc}%
A{\mathbb{I}}_{2} & C\sigma_{z}\\
C\sigma_{z} & B{\mathbb{I}}_{2}%
\end{array}
\right)  \label{CM_2}%
\end{equation}
where%
\begin{align}
A  &  =\cosh(2r)+2\bar{n}_{1}\cosh^{2}r+2\bar{n}_{2}\sinh^{2}r\nonumber\\
B  &  =\cosh(2r)+2\bar{n}_{1}\sinh^{2}r+2\bar{n}_{2}\cosh^{2}r\nonumber\\
C  &  =(1+\bar{n}_{1}+\bar{n}_{2})\sinh2r, \label{ABCterms}%
\end{align}
with ${\mathbb{I}}_{2}$ the $2\times2$ identity matrix and $\sigma
_{z}=\text{diag}(1,-1)$ the z-Pauli matrix. This corresponds to considering a
density operator of the form
\[
\rho=S_{2}(r)\left(  \nu_{1}\otimes\nu_{2}\right)  S_{2}(r)^{\dagger}\,,
\]
where $S_{2}(r)=\exp\{r(a^{\dagger}b^{\dagger}-ab)\}$ is the two-mode
squeezing operator. This state depends on three real parameters: the squeezing
parameter and the two thermal numbers, i.e., we have
\[
\rho=\rho(r,\bar{n}_{1},\bar{n}_{2})~.
\]
In particular, for $\bar{n}_{1}=\bar{n}_{2}=0$ the state is pure and
corresponds to a two-mode squeezed vacuum state $\rho(r,0,0)=S_{2}%
(r)(\left\vert 0\right\rangle _{1}\left\langle 0\right\vert \otimes\left\vert
0\right\rangle _{2}\left\langle 0\right\vert )S_{2}^{\dagger}(r)$.
%%%

\section{Quantum Chernoff bound for Gaussian states}

\label{s:qcbG} Here we review the formula of the QCB for multimode Gaussian
states~\cite{Pirandola}. In particular, we adapt this formula to our notation
and physical units (here the vacuum noise is $1/2$, while in
Ref.~\cite{Pirandola} it was equal to $1$). Let us consider two Gaussian
states $\rho$ (with statistical moments $\boldsymbol{X}$ and
$\boldsymbol{\sigma}$) and $\rho^{\prime}$ (with statistical moments
$\boldsymbol{X}^{\prime}$ and $\boldsymbol{\sigma}^{\prime}$). The CMs of
these two states can be decomposed as%
\begin{align}
\boldsymbol{\sigma}  &  =\boldsymbol{S~\boldsymbol{W}}(\bar{n}_{1},\cdots
,\bar{n}_{n})~\boldsymbol{S^{T}}\label{deco1}\\
\boldsymbol{\sigma}^{\prime}  &  =\boldsymbol{S}^{\prime}%
~\boldsymbol{\boldsymbol{W}}(\bar{n}_{1}^{\prime},\cdots,\bar{n}_{n}^{\prime
})~\boldsymbol{S}^{\prime}\boldsymbol{^{T}}, \label{deco2}%
\end{align}
where $\{\bar{n}_{k}\}$ and $\{\bar{n}_{k}^{\prime}\}$ are their thermal
numbers, and%
\[
\boldsymbol{W}(x_{1},\cdots,x_{n})=\bigoplus_{k=1}^{n}(2x_{k}+1){\mathbb{I}%
}_{2}~.
\]
Now let us define the functions%
\[
G_{s}(x)=\frac{1}{(x+1)^{s}-x^{s}}~,
\]
and
\[
\Lambda_{s}(x)=\frac{x^{s}}{(x+1)^{s}-x^{s}}~.
\]
Then, the QCB is given by
\[
Q=\inf_{0\leq s\leq1}Q_{s}~,
\]
where%
\begin{equation}
Q_{s}=\frac{\Pi_{s}}{\sqrt{\det\boldsymbol{\Sigma}_{s}}}\exp\left(  -\frac
{1}{2}\mathbf{d}^{T}\boldsymbol{\Sigma}_{s}^{-1}\mathbf{d}\right)  ~.
\label{Qesse}%
\end{equation}
In the formula of Eq.~(\ref{Qesse}), we have $\mathbf{d}=\boldsymbol{X}%
-\boldsymbol{X}^{\prime}$,
\[
\Pi_{s}=\prod_{k=1}^{n}G_{s}(\bar{n}_{k})G_{1-s}(\bar{n}_{k}^{\prime})~,
\]
and%
\begin{align*}
\boldsymbol{\Sigma}_{s} =  &  \boldsymbol{S~W}[\Lambda_{s}(\bar{n}_{1}%
),\cdots,\Lambda_{s}(\bar{n}_{n})]~\boldsymbol{S^{T}}\\
&  +\boldsymbol{S}^{\prime}~\boldsymbol{W}[\Lambda_{1-s}(\bar{n}_{1}^{\prime
}),\cdots,\Lambda_{1-s}(\bar{n}_{n}^{\prime})]~\boldsymbol{S}^{\prime T}~.
\end{align*}

\subsection{Discrimination of squeezed thermal states}

For the discrimination of squeezed thermal states, the previous formula
simplifies a lot. First of all, since they are zero-mean Gaussian states, we
have $\mathbf{d}=0$ and, therefore, the exponential factor in Eq.~(\ref{Qesse}%
) disappears. Then, the symplectic decompositions in Eqs.~(\ref{deco1})
and~(\ref{deco2}) are achieved using symplectic matrices $\boldsymbol{S}$ and
$\boldsymbol{S}^{\prime}$ which are just one-parameter squeezing matrices,
i.e., $\boldsymbol{S}=\boldsymbol{S}(r)$ and $\boldsymbol{S}^{\prime
}=\boldsymbol{S}^{\prime}(r^{\prime})$.

Thus, let us consider the discrimination of single-mode squeezed thermal
states $\rho=\rho(r,\bar{n})$ and $\rho^{\prime}=\rho^{\prime}(r^{\prime}%
,\bar{n}^{\prime})$. In this case, the QCB can be computed using%
\begin{equation}
Q_{s}=\frac{\Pi_{s}(\bar{n},\bar{n}^{\prime})}{\sqrt{\det\boldsymbol{\Sigma
}_{s}(r,\bar{n},r^{\prime},\bar{n}^{\prime})}}~, \label{Qs1mode}%
\end{equation}
where
\[
\Pi_{s}(\bar{n},\bar{n}^{\prime})=G_{s}(\bar{n})G_{1-s}(\bar{n}^{\prime})~,
\]
and%
\begin{align*}
\boldsymbol{\Sigma}_{s}(\bar{n},\bar{n}^{\prime},r,r^{\prime})  &
=\boldsymbol{S}(r)\boldsymbol{W}[\Lambda_{s}(\bar{n})]\boldsymbol{S}%
(r)\boldsymbol{^{T}}\\
&  +\boldsymbol{S}(r^{\prime})\boldsymbol{W}[\Lambda_{1-s}(\bar{n}^{\prime
})]\boldsymbol{S}(r^{\prime})\boldsymbol{^{T}}.
\end{align*}

For the discrimination of two-mode squeezed thermal states $\rho=\rho
(r,\bar{n}_{1},\bar{n}_{2})$ and $\rho^{\prime}=\rho^{\prime}(r^{\prime}%
,\bar{n}_{1}^{\prime},\bar{n}_{2}^{\prime})$, we can use
\begin{equation}
Q_{s}=\frac{\Pi_{s}(\bar{n}_{1},\bar{n}_{2},\bar{n}_{1}^{\prime},\bar{n}%
_{2}^{\prime})}{\sqrt{\det\boldsymbol{\Sigma}_{s}(r,\bar{n}_{1},\bar{n}%
_{2},r^{\prime},\bar{n}_{1}^{\prime},\bar{n}_{2}^{\prime})}}~, \label{Qs2mode}%
\end{equation}
where%
\[
\Pi_{s}(\bar{n}_{1},\bar{n}_{2},\bar{n}_{1}^{\prime},\bar{n}_{2}^{\prime
})=G_{s}(\bar{n}_{1})G_{s}(\bar{n}_{2})G_{1-s}(\bar{n}_{1}^{\prime}%
)G_{1-s}(\bar{n}_{2}^{\prime})~,
\]
and%
\begin{gather*}
\boldsymbol{\Sigma}_{s}(r,\bar{n}_{1},\bar{n}_{2},r^{\prime},\bar{n}%
_{1}^{\prime},\bar{n}_{2}^{\prime})=\boldsymbol{S}(r)\boldsymbol{W}%
[\Lambda_{s}(\bar{n}_{1}),\Lambda_{s}(\bar{n}_{2})]\boldsymbol{S}%
(r)\boldsymbol{^{T}}\\
+\boldsymbol{S}(r^{\prime})\boldsymbol{W}[\Lambda_{1-s}(\bar{n}_{1}^{\prime
}),\Lambda_{1-s}(\bar{n}_{2}^{\prime})]\boldsymbol{S}(r^{\prime}%
)\boldsymbol{^{T}}.
\end{gather*}

\section{Detection of losses by thermal probes}

\label{QCBloss} In what follows, we study the evolution of a Gaussian state in
a dissipative channel $\mathcal{E}_{\Gamma}$ characterized by a damping rate
$\Gamma$, which may result from the interaction of the system with an external
environment, as for example a bath of oscillators, or from an absorption
process. We consider the problem of detecting whether or not the dissipation
dynamics occurred. Given an input state $\rho$, this corresponds to
discriminating between an output state identical to the input $\rho$, and
another output state storing the presence of loss $\mathcal{E}_{\Gamma}(\rho
)$. Lossy channels are Gaussian channels, meaning that they tansform Gaussian
states into Gaussian states. Furthermore, if the input is a squeezed thermal
state, then the output state is still squeezed thermal (this is discussed in
detail afterwards).

\begin{figure}[ptb]
\vspace{-1.0cm}
\par
\begin{center}
\includegraphics[width=0.5\textwidth]{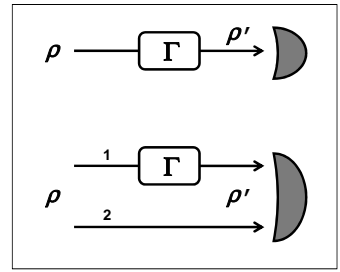}
\end{center}
\par
\vspace{-1.1cm} \caption{Single- and two-mode schemes for the detection of
losses. Top: a single-mode squeezed thermal state $\rho$ enters the lossy
channel with damping rate $\Gamma$. A measurement apparatus detects the output
state $\rho^{\prime}$. Bottom: the lossy channel acts on the probing mode (1)
of a two-mode squeezed thermal state $\rho$, while the reference mode (2)
bypasses the channel. The output state $\rho^{\prime}$ of both the modes is
then measured.}%
\label{setup1e2}%
\end{figure}

In general, we consider the schematic diagram depicted in Fig.\ref{setup1e2}.
In order to detect loss, we consider either a single-mode squeezed thermal
state evolving in the lossy channel with parameter $\Gamma$ followed by a
measurement at the output, or a two-mode squeezed thermal state with the
damping process occurring in only one of the two modes (the probing mode),
followed by a measurement on both of the modes. Our aim is to minimize the
error probability in discriminating between the ideal case $\Gamma=0$ and the
lossy case $\Gamma>0$. In the next section, this will be done by fixing some
important parameters of the input state, such as total energy and
squeezing~\cite{ContQCB}. 

The propagation of a mode of radiation in a lossy channel corresponds to the
coupling of the mode $a$ with a zero temperature reservoir made of large
number of external modes. By assuming a Markovian reservoir and weak coupling
between the system and the reservoir the dynamics of the system is described
by the Lindblad Master equation \cite{Walls}
\begin{equation}
\dot{\rho}=\frac{\Gamma}{2}\mathcal{L}[a]\rho\label{ME}%
\end{equation}
where $\mathcal{L}[a]\rho=2a\rho a^{\dagger}-a^{\dagger}a\rho-\rho a^{\dagger
}a$. The general solution may be expressed by using the operator-sum
representation of the associated completely-positive map, i.e., upon writing
$\eta=e^{-\Gamma t}$, we have%
\[
\varrho(\eta)=\sum_{m}V_{m}\varrho V_{m}^{\dag}%
\]
where
\[
V_{m}=\sqrt{\frac{(1-\eta)^{m}}{m!}}a^{m}\eta^{\frac{1}{2}(a^{\dag}a-m)}\,,
\]
and $\varrho$ is the initial state.

\subsection{Single-mode case}

Let us now start with single-mode states. Eq. (\ref{ME}) can be recast into a
Fokker-Planck equation for the Wigner function in terms of the quadrature
variables $q$ and $p$,
\[
\dot{W}=\frac{\Gamma}{2}\left[  \partial_{\boldsymbol{X}}^{T}\boldsymbol{X}%
+\partial_{\boldsymbol{X}}^{T}\boldsymbol{\sigma}_{\infty}\partial
_{\boldsymbol{X}}^{T}\right]  W
\]
where $\boldsymbol{X}=(q,p)^{T}$, $\partial_{\boldsymbol{X}}=(\partial
_{q},\partial_{p})^{T}$ and we introduced the diffusion matrix
$\boldsymbol{\sigma}_{\infty}=\text{diag}(1/2,1/2)$. Solving the equation for
the Wigner function of a single-mode Gaussian state one can obtain the
evolution equation for the CM $\boldsymbol{\sigma}$. This is given by
\cite{AS1}
\[
\dot{\boldsymbol{\sigma}}=-\Gamma(\boldsymbol{\sigma}-\boldsymbol{\sigma
}_{\infty})~,
\]
which yields to
\[
\boldsymbol{\sigma}(t)=e^{-\Gamma t}\boldsymbol{\sigma}_{0}+(1-e^{-\Gamma
t}){\boldsymbol{\sigma}}_{\infty}~.
\]
The latter equation describes the evolution of an initial Gaussian state with
CM $\boldsymbol{\sigma}_{0}$ towards the stationary state given by the
Gaussian state of the environment with CM $\boldsymbol{\sigma}_{\infty}$. For
simplicity, from now on we omit the index of $\boldsymbol{\sigma}_{0}$, we
replace $\boldsymbol{\sigma}(t)\rightarrow\boldsymbol{\sigma}^{\prime}$, and
we insert the time $t$ into the damping parameter $\Gamma$. Thus, the evolved
CM of the single mode case simply reads
\[
\boldsymbol{\sigma}^{\prime}=e^{-\Gamma}\boldsymbol{\sigma}+(1-e^{-\Gamma
})\,\boldsymbol{\sigma}_{\infty}~.
\]

Now let us consider the specific case of an input squeezed thermal state
$\rho=\rho(r,n_{T})$ with squeezing $r$\ and thermal number $n_{T}$. According
to Eqs.~(\ref{CM_1}) and~(\ref{ExpressionsSIMPLE}), its CM is given by
\[
\boldsymbol{\sigma}=\left(
\begin{array}
[c]{cc}%
a & 0\\
0 & b
\end{array}
\right)  ~,
\]
with%
\begin{equation}
a=\frac{1}{2}\left(  1+2n_{\scriptstyle T}\right)  e^{2r}~,~b=\frac{1}%
{2}\left(  1+2n_{\scriptstyle T}\right)  e^{-2r}\,~.
\end{equation}
At the output of the channel, the state $\rho^{\prime}$ has CM
\[
\boldsymbol{\sigma}^{\prime}=\left(
\begin{array}
[c]{cc}%
a_{\Gamma} & 0\\
0 & b_{\Gamma}%
\end{array}
\right)  ~,
\]
where
\begin{equation}
a_{\Gamma}=\frac{1}{2}\left(  1+2n_{\Gamma}\right)  e^{2r_{\Gamma}%
}~,~b_{\Gamma}=\frac{1}{2}\left(  1+2n_{\Gamma}\right)  e^{-2r_{\Gamma}}~,
\end{equation}
and
\begin{align}
n_{\Gamma}  &  =\sqrt{\det[\boldsymbol{\sigma}^{\prime}]}-1/2\\
r_{\Gamma}  &  =\frac{1}{4}\log\left[  \frac{e^{-\Gamma}a+(1-e^{-\Gamma}%
)/2}{e^{-\Gamma}b+(1-e^{-\Gamma})/2}\right]  .
\end{align}
Thus, we still have a squeezed thermal state $\rho^{\prime}=\rho^{\prime
}(r_{\Gamma},n_{\Gamma})$ with squeezing $r_{\Gamma}$\ and thermal number
$n_{\Gamma}$. Now, the discrimination between a lossless ($\Gamma=0$) and a
lossy channel ($\Gamma>0$) corresponds to the discrimination between the input
state $\rho=\rho(r,n_{T})$ and the output one $\rho^{\prime}=\rho^{\prime
}(r_{\Gamma},n_{\Gamma})$. In order to estimate the error probability
affecting this discrimination, we can compute the quantum Chernoff bound. This
is achieved by replacing
\[
(r,\bar{n})\rightarrow(r,n_{T})\;\; \hbox{and}\;\;(r^{\prime},\bar{n}^{\prime
})\rightarrow(r_{\Gamma},n_{\Gamma})\,
\]
in Eq.~(\ref{Qs1mode}).

\subsection{Two-mode case}

According to the scheme of Fig.~\ref{setup1e2}, the map describing the
evolution of a two-mode state is $\mathcal{E}_{\Gamma}\otimes\mathcal{I}$,
where the lossy channel $\mathcal{E}_{\Gamma}$\ acts on the probing mode while
the identity channel $\mathcal{I}$\ acts on the reference mode. At the level
of the CM it corresponds to the following transformation%
\begin{align}
\boldsymbol{\sigma}^{\prime} =  &  \left(  e^{-\Gamma/2}{\mathbb{I}}_{2}%
\oplus{\mathbb{I}}_{2}\right)  \boldsymbol{\sigma}\left(  e^{-\Gamma
/2}{\mathbb{I}}_{2}\oplus{\mathbb{I}}_{2}\right) \nonumber\\
&  +({\mathbb{I}}_{4}-e^{-\Gamma}{\mathbb{I}}_{2}\oplus{\mathbb{I}}%
_{2})\boldsymbol{\sigma}_{\infty}~. \label{IOrelation}%
\end{align}
As input state, let us consider a two-mode squeezed thermal state $\rho
=\rho(r,n_{\scriptstyle T_{1}},n_{\scriptstyle T_{2}})$. Its CM is provided in
Eq.~(\ref{CM_2}) with the elements given in Eq.~(\ref{ABCterms}) by replacing
$(r,\bar{n}_{1},\bar{n}_{2})\rightarrow(r,n_{\scriptstyle T_{1}}%
,n_{\scriptstyle T_{2}})$. The CM\ of the output state can be derived using
the Eq.~(\ref{IOrelation}). This CM can be put in the normal form of
Eq.~(\ref{CM_2}) with elements given by Eq.~(\ref{ABCterms}) by replacing
$(r,\bar{n}_{1},\bar{n}_{2})\rightarrow(r_{\Gamma},n_{\scriptstyle\Gamma_{1}%
},n_{\scriptstyle\Gamma_{2}})$. Here the squeezing parameter $r_{\Gamma}$ and
the thermal numbers $n_{\scriptstyle\Gamma_{1}}$ and $n_{\scriptstyle\Gamma
_{2}}$ are function of the input parameters $r$, $n_{\scriptstyle T_{1}}$ and
$n_{\scriptstyle T_{2}}$ (the explicit expression is too long to be shown
here). Thus, the output state is still a two-mode squeezed thermal state
$\rho^{\prime}=\rho^{\prime}(r_{\Gamma},n_{\scriptstyle\Gamma_{1}%
},n_{\scriptstyle\Gamma_{2}})$.

As before, the discrimination between a lossless ($\Gamma=0$) and a lossy
channel ($\Gamma>0$) corresponds to the discrimination between the input state
$\rho=\rho(r,n_{\scriptstyle T_{1}},n_{\scriptstyle T_{2}})$ and the output
one $\rho^{\prime}=\rho^{\prime}(r_{\Gamma},n_{\scriptstyle\Gamma_{1}%
},n_{\scriptstyle\Gamma_{2}})$. The error probability affecting this
discrimination is estimated by the QCB which is computed by replacing
$(r,\bar{n}_{1},\bar{n}_{2})\rightarrow(r,n_{\scriptstyle T_{1}}%
,n_{\scriptstyle T_{2}})$ and $(r^{\prime},\bar{n}_{1}^{\prime},\bar{n}%
_{2}^{\prime})\rightarrow(r_{\Gamma},n_{\scriptstyle\Gamma_{1}}%
,n_{\scriptstyle\Gamma_{2}})$ in Eq.~(\ref{Qs2mode}).

\section{Optimization of the thermal probes}

\label{results} In this section, we optimize the discrimination of a lossless
($\Gamma=0$) from a lossy channel ($\Gamma>0$) by maximizing over thermal
probes, i.e., single- and two-mode squeezed thermal states. For this sake, we
evaluate the QCB as a function of the most important parameters of the input
state, i.e., its total energy and squeezing. In our first analysis, we show
that for fixed total energy, single- and two-mode squeezed vacuum states are
optimal. In particular, we show the conditions where the two-mode state
outperforms the single-mode counterpart. Then, by fixing \textit{both} the
total energy and squeezing, we will find the optimal squeezed thermal state.
According to Sec.~\ref{QHT} the minimization of the QCB over single-copy
states implies the minimization over multi-copy states (when the minimization
is unconstrained or subject to single-copy constraints). This implies that
finding the optimal input state $\rho$ at fixed energy automatically assures
that $\rho\otimes\rho\otimes\cdots$ is the optimal multi-copy state at fixed
energy\textit{ per copy} when we consider a multiple access to the unknown
(memoryless) channel.

In order to perform our investigation we introduce a suitable parametrization
of the input energy. Given a single-mode squeezed thermal state $\rho
=\rho(r,n_{T})$, its energy (mean total number of photons) can be written as%
\begin{equation}
N_{1}=n_{\scriptstyle T}+n_{\scriptstyle S}+2n_{\scriptstyle S}%
n_{\scriptstyle T}~, \label{nt1}%
\end{equation}
where $n_{\scriptstyle T}$ accounts for the mean number of thermal photons,
$n_{\scriptstyle S}=\sinh^{2}r$ quantifies the squeezing, and
$n_{\scriptstyle
S}n_{\scriptstyle T}$ is a cross term. Alternatively, we can introduce a
squeezing factor $\beta_{1}\in\lbrack0,1]$ such that%
\begin{align}
n_{\scriptstyle S}  &  =\beta_{1}N_{1}\\
n_{\scriptstyle T}  &  =\frac{(1-\beta_{1})N_{1}}{1+2\beta_{1}N_{1}}~.
\end{align}
Thus the single-mode squeezed thermal state can be parametrized as $\rho
=\rho(N_{1},\beta_{1})$, i.e., in terms of its total energy $N_{1}$ and the
squeezing factor $\beta_{1}$. Note that for $\beta_{1}=0$ the state is
completely thermal with energy $N_{1}=n_{\scriptstyle T}$, while for
$\beta_{1}=1$ the state is a squeezed vacuum with energy $N_{1}%
=n_{\scriptstyle S}$. In our problem of loss detection ($\Gamma=0$ versus
$\Gamma>0$), we denote by $Q_{1}(N_{1},\beta_{1})$ the output QCB which is
computed by using the input state $\rho=\rho(N_{1},\beta_{1})$.

Now, given a two-mode squeezed thermal state $\rho=\rho(r,n_{\scriptstyle
T_{1}},n_{\scriptstyle T_{2}})$, its total energy can be written as%
\begin{equation}
N_{2}=n_{\scriptstyle T_{1}}+n_{\scriptstyle T_{2}}+2n_{\scriptstyle S}%
+2n_{\scriptstyle S}(n_{\scriptstyle T_{1}}+n_{\scriptstyle T_{2}})~,
\label{nt2}%
\end{equation}
where $n_{\scriptstyle T_{1}}$ ($n_{\scriptstyle T_{2}}$)\ quantifies the
thermal photons in the probing (reference) mode, $n_{\scriptstyle S}=\sinh
^{2}r$ quantifies the two-mode squeezing energy, and the last energetic term
is a cross term. In this case, besides the squeezing factor $\beta_{2}$, we
can also introduce an asymmetry parameter $\gamma\in\lbrack0,1]$ which
quantifies the fraction of thermal energy used for the probing mode. In other
words, we can write
\begin{align}
n_{\scriptstyle S}  &  =\frac12 \beta_{2} N_{2}\\
n_{\scriptstyle T_{1}}  &  =\gamma\,\frac{(1-\beta_{2})N_{2}}{1+\beta_{2}%
N_{2}}\\
n_{\scriptstyle T_{2}}  &  =(1-\gamma)\frac{(1-\beta_{2})N_{2}}{1+\beta
_{2}N_{2}}.
\end{align}
Thus the two-mode squeezed thermal state can be parametrized as $\rho
=\rho(N_{2},\beta_{2},\gamma)$, i.e., in terms of the total energy $N_{2}$,
the squeezing factor $\beta_{2}$, and the asymmetry parameter $\gamma$. Note
that for $\beta_{2}=0$ we have two thermal states, one describing the probing
mode with thermal energy $n_{\scriptstyle T_{1}}=\gamma\,N_{2}$, and the other
one describing the reference mode with thermal energy $n_{\scriptstyle T_{2}%
}=(1-\gamma)N_{2}$. For $\beta_{2}=1$ we have instead a two-mode squeezed
vacuum state with total energy $N_{2}=2n_{\scriptstyle S}$. In this case, the
thermal energy is zero and $\gamma$ can be therefore arbitrary. In our problem
of loss detection ($\Gamma=0$ versus $\Gamma>0$), we denote by $Q_{2}%
(N_{2},\beta_{2},\gamma)$ the output QCB which is computed by using the input
state $\rho=\rho(N_{2},\beta_{2},\gamma)$.
%%%%

\subsection{Optimal input at fixed total energy}

In our first investigation we fix the mean total number of photons of the
input state. In other words we fix%
\begin{equation}
N_{1}=N_{2}=N~.
\end{equation}
Then we minimize the output QCB among single-mode and two-mode squeezed
thermal states. As a first step we compute the optimal quantities%
\begin{align}
Q_{1}(N)  &  :=\inf_{\beta_{1}}Q_{1}(N,\beta_{1})\label{qcb1}\\
Q_{2}(N)  &  :=\inf_{\beta_{2},\gamma}Q_{2}(N,\beta_{2},\gamma)~. \label{qcb2}%
\end{align}
Then, we compare $Q_{1}(N)$ with $Q_{2}(N)$.\begin{figure}[h]
\includegraphics[width=0.95\columnwidth]{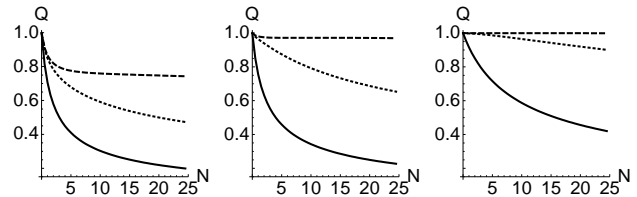} \caption{Output
QCB$\ Q_{1}(N,\beta_{1})$ optimized over input single-mode squeezed thermal
states $\rho=\rho(N,\beta_{1})$. From left to right we consider different
values of the transmissivity: $\eta=0.1$ (left panel), $\eta=0.5$ (middle
panel) and $\eta=0.9$ (right panel). In each panel, we plot $Q_{1}(N,\beta
_{1})$ as function of the energy $N$\ for different values of $\beta_{1}$.
From top to bottom: $\beta_{1}=0.1$ (dashed line), $\beta_{1}=0.5$ (dotted
line) and $\beta_{1}=1$ (solid line). The minimum curve is always achieved for
$\beta_{1}=1$, i.e., for an input single-mode squeezed vacuum state.}%
\label{b1fig}%
\end{figure}\begin{figure}[h]
\includegraphics[width=0.92\columnwidth]{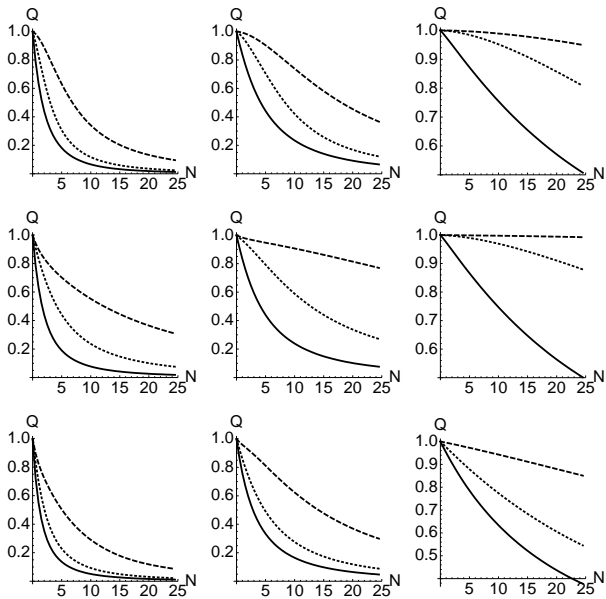} \caption{Output
QCB$\ Q_{2}(N,\beta_{2},\gamma)$ optimized over input two-mode squeezed
thermal states $\rho=\rho(N,\beta_{2},\gamma)$. From left to right we consider
different values of the transmissivity: $\eta=0.1,$ $0.5$ and $0.9$. From top
to bottom, we consider different values of the asymmetry parameter $\gamma=0,$
$0.5$ and $1$. In each panel, we then plot $Q_{2}$ as function of the energy
$N$\ for different values of $\beta_{2}$. From top to bottom: $\beta_{2}=0.1$
(dashed line), $\beta_{2}=0.5$ (dotted line) and $\beta_{2}=1$ (solid line).
The minimum curve is always achieved for $\beta_{2}=1$ corresponding to an
input two-mode squeezed vacuum state.}%
\label{b2fig}%
\end{figure}
%%%

\begin{figure}[h]
\includegraphics[width=0.7\columnwidth]{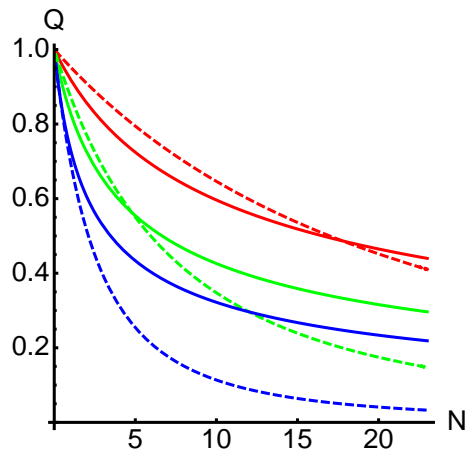}\caption{(Color
online). Single-mode QCB $Q_{1}(N)$ (solid lines) and two-mode QCB $Q_{2}%
(N)$\ (dashed lines) as a function of the input energy $N$ for different
damping rates. From top to bottom $\Gamma=0.1,0.3,1$ (red, green and blue,
respectively) corresponding to $\eta\simeq0.9,0.74,0.37$. By comparing curves
with the same color (fixed damping $\Gamma$), we can see that $Q_{2}(N)$
outperforms $Q_{1}(N)$ only after a certain value of the input energy $N$.}%
\label{beta1NG}%
\end{figure}

According to our findings, in the Eqs.~(\ref{qcb1}) and~(\ref{qcb2}) the
infima are achieved for $\beta_{1}=\beta_{2}=1$. This is numerically shown in
Fig.~\ref{b1fig} for the single-mode case and in Fig.~\ref{b2fig} for the
two-mode case. Thus, we have found that, at fixed input energy $N$, the
optimal thermal probes are given by single- and two-mode squeezed vacuum
states. In this case, the input state is pure and the QCB corresponds to the
fidelity (which is the case when the s-overlap in Eq. (\ref{QCBgen}) is
minimized for $s$ approching the border). Let us adopt the transmissivity
$\eta=e^{-\Gamma}$ to quantify the damping of the channel, so that $\Gamma=0$
(ideal channel) corresponds to $\eta=1$, and $\Gamma>0$ (lossy channel)
corresponds to $0\leq\eta<1$. Then, for single-mode we can write%
\begin{equation}
Q_{1}(N)=\frac{1}{\sqrt{1+N(1-\eta^{2})}}~, \label{QCB_1mode_ANA}%
\end{equation}
and for two-modes we derive%
\begin{equation}
Q_{2}(N)=\frac{4}{\left[  2+N(1-\sqrt{\eta})\right]  ^{2}}~.
\label{QCB_2mode_ANA}%
\end{equation}
In Fig. \ref{beta1NG}, we show the behaviors of the single-mode QCB $Q_{1}(N)$
and two-mode QCB $Q_{2}(N)$ as function of the input energy $N$\ for several
values of transmissivity $\eta$ (or, equivalently, the damping rate $\Gamma$).
As expected the discrimination improves by increasing the input energy $N$ and
decreasing the transmissivity $\eta$.
%%%%%%%%
%%%%%%%
\begin{figure}[h]
\includegraphics[width=0.78\columnwidth]{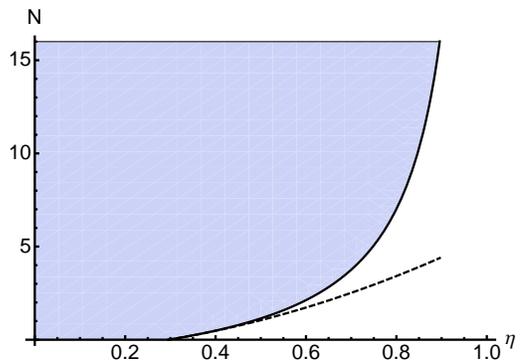} \caption{Threshold
energy as a function of the transmissivity $N_{th}=N_{th}(\eta)$ (solid curve
dividing the dark and the white areas). The dark area indicates the values of
the energy $N$ for which the two-mode squeezed vacuum state is optimal. The
white region indicates where the single-mode squeezed vacuum state is optimal.
The dashed line denotes the behavior of the threshold energy $N_{th}$ close to
the critical transmissivity $\eta_{c}\simeq0.296$.}%
\label{f:f3_th}%
\end{figure}
%%%%%%%

As we can see from Fig.~\ref{beta1NG}, for a given value of the transmissivity
$\eta$, the two-mode QCB $Q_{2}(N)$ outperforms the single-mode QCB $Q_{1}%
(N)$\ only after a threshold energy. In fact, for any value of the
transmissivity $\eta$\ larger than a critical value $\eta_{c}$ there is a
threshold energy $N_{th}=N_{th}(\eta)$ that makes the two-mode squeezed vacuum
state more convenient than the single-mode counterpart. This threshold energy
decreases for decreasing values of $\eta$. In particular, for transmissivities
less than the critical value $\eta_{c}$, the threshold energy becomes zero,
i.e., the two-mode state is always better than single-mode state. We have
numerically evaluated the critical value $\eta_{c}\simeq0.296$ (corresponding
to $\Gamma_{c}\simeq1.22$). This phenomenon is fully illustrated in
Fig.~\ref{f:f3_th}, where we have plotted the threshold energy as function of
the transmissivity $N_{th}=N_{th}(\eta)$. For $N>N_{th}$ (dark area), the
optimal state is the two-mode squeezed vacuum state, while for $N<N_{th}$
(white area) it is the single-mode squeezed vacuum state. In particular, note
that $N_{th}=0$ at $\eta=\eta_{c}$. Close to the critical transmissivity we
have~\cite{OtherCurve}
\begin{equation}
N_{th}\simeq4(\eta-\eta_{c})+5.5(\eta-\eta_{c})^{2}~.
\end{equation}

\subsection{Optimal input at fixed energy and squeezing}

It should be said that, in realistic conditions, it is unlikely to have pure
squeezing. For this reason, it is important to investigate the performances of
the squeezed thermal states by fixing this physical parameter together with
the total energy. Thus, in this section, we fix both the input energy and
squeezing, i.e., we set%
\begin{align}
N_{1}  &  =N_{2}=N,\nonumber\\
\beta_{1}  &  =\beta_{2}=\beta~~(0\leq\beta\leq1)~.
\end{align}
Then, we compare the single-mode squeezed thermal state $\rho=\rho(N,\beta)$
with the two-mode squeezed thermal states $\rho=\rho(N,\beta,\gamma)$ for
various values of $\gamma$. In other words, we compare $Q_{1}(N,\beta)$ and
$Q_{2}(N,\beta,\gamma)$.

\begin{figure}[h]
\includegraphics[width=0.48\columnwidth]{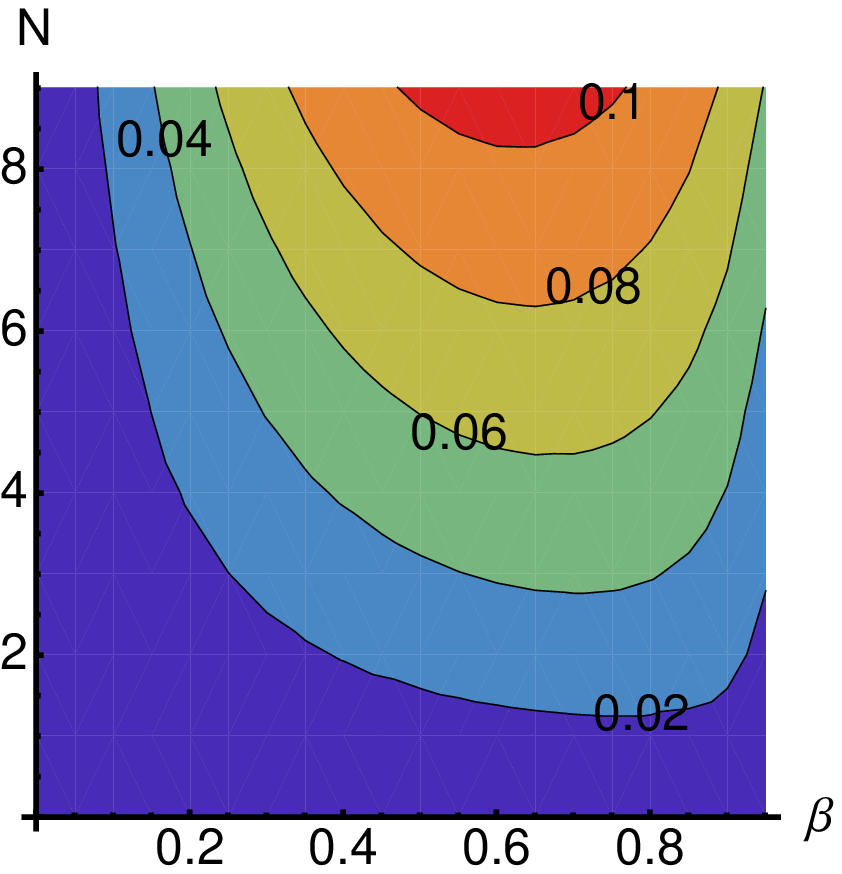}
\includegraphics[width=0.48\columnwidth]{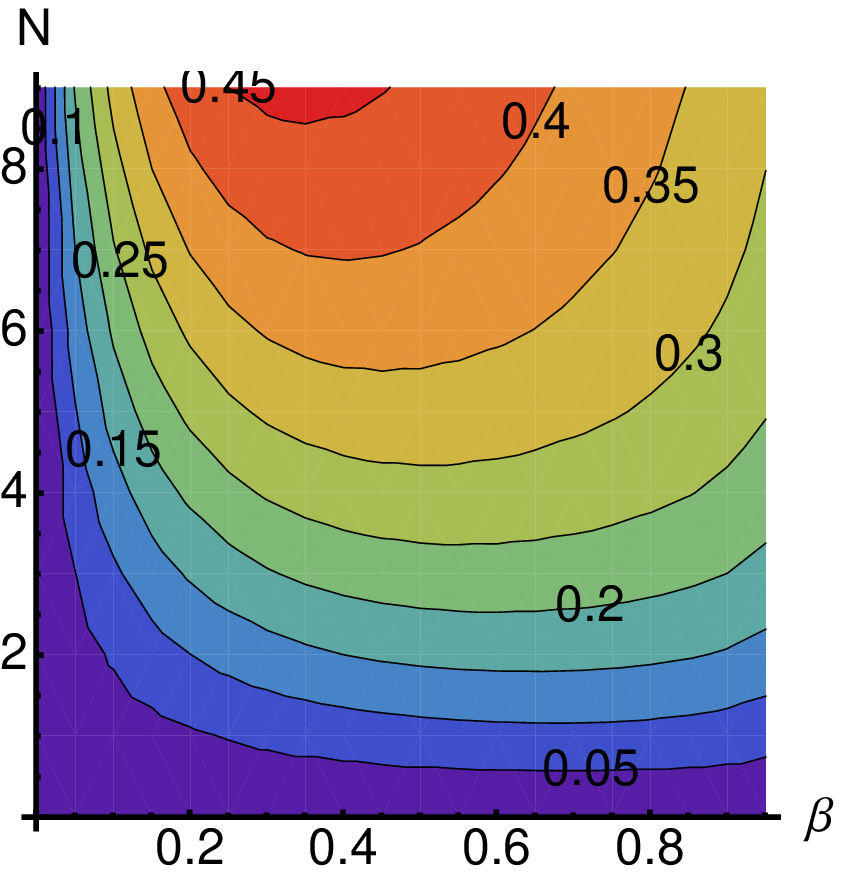}\caption{(Color online)
Density plot of the QCB reduction $\Delta Q$ as function of the input energy
$N$\ and the squeezing $\beta$. The left plot is for $\Gamma=0.1$ and the
right one for $\Gamma=0.9$.}%
\label{f:f4_dq}%
\end{figure}

For fixed $N$ and $\beta$, we find that the minimum of $Q_{2}(N,\beta,\gamma)$
is achieved for $\gamma=1$ (easy to check numerically). This means that
two-mode discrimination is easier when all the thermal photons are sent
through the lossy channel. In this case we find numerically that
\[
Q_{2}(N,\beta,1)<Q_{1}(N,\beta)\,,
\]
for every values of the input parameters $N$ and $\beta$, and every value of
damping rate $\Gamma$ in the channel. In other words, at fixed energy and
squeezing, there is a two-mode squeezed thermal state (the asymmetric one with
$\gamma=1$) able to outperform the single-mode squeezed thermal state in the
detection of any loss. In order to quantify the improvement we introduce the
QCB reduction
\[
\Delta Q=Q_{1}(N,\beta)-Q_{2}(N,\beta,1).
\]
The more positive this quantity is, the more convenient is the use of the
two-mode state instead of the single-mode one. In Fig. \ref{f:f4_dq} we show
the behavior of $\Delta Q$ as function of the input energy and squeezing for
two different values of the damping. As one can see from the plot, the
QCB\ reduction is always positive. Its value increases with the energy while
reaching a maximum for intermediate values of the squeezing. By comparing the
two panels of Fig.~\ref{f:f4_dq}, we can also note that the QCB reduction
increases for increasing damping $\Gamma$ (i.e., decreasing transmissivity).

\begin{figure}[h]
\includegraphics[width=0.48\columnwidth]{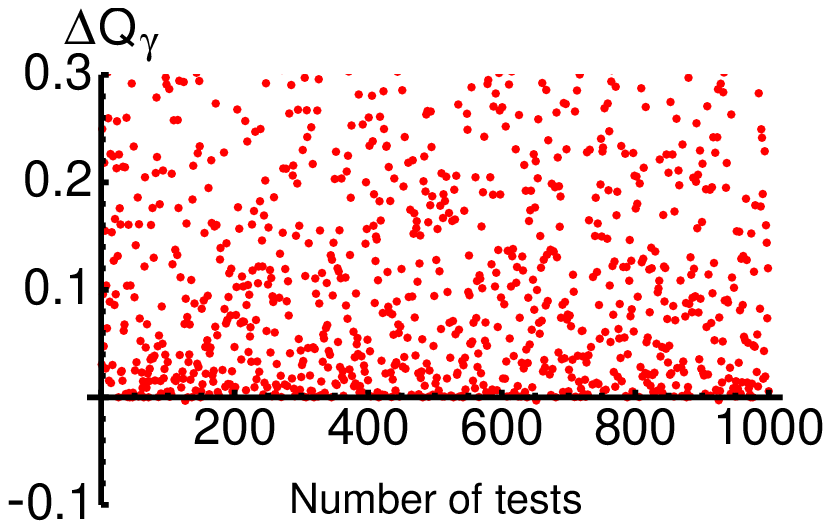}
\includegraphics[width=0.48\columnwidth]{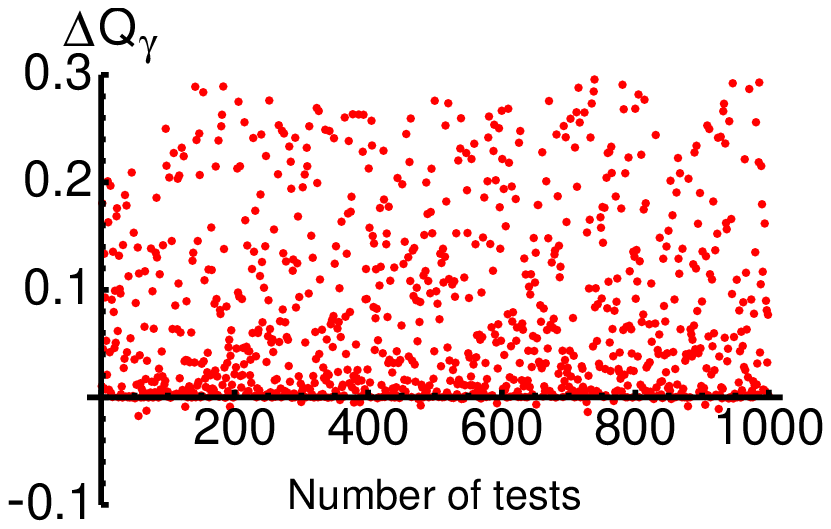}
\includegraphics[width=0.48\columnwidth]{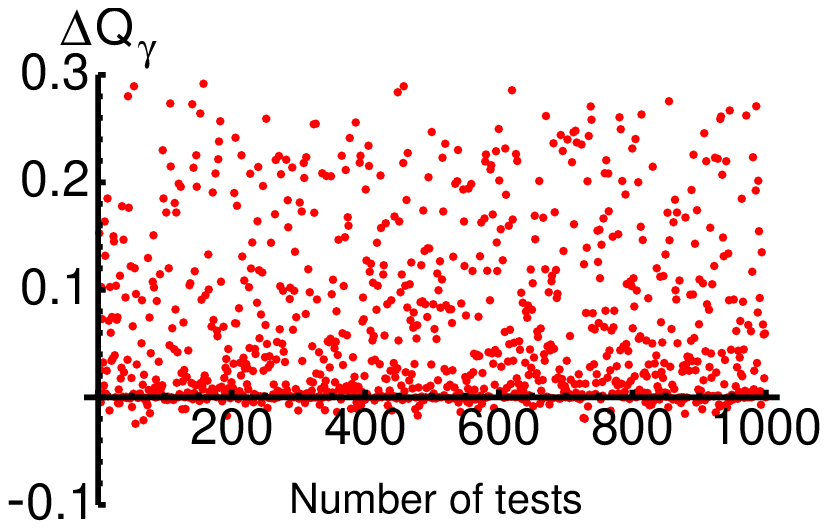}
\includegraphics[width=0.48\columnwidth]{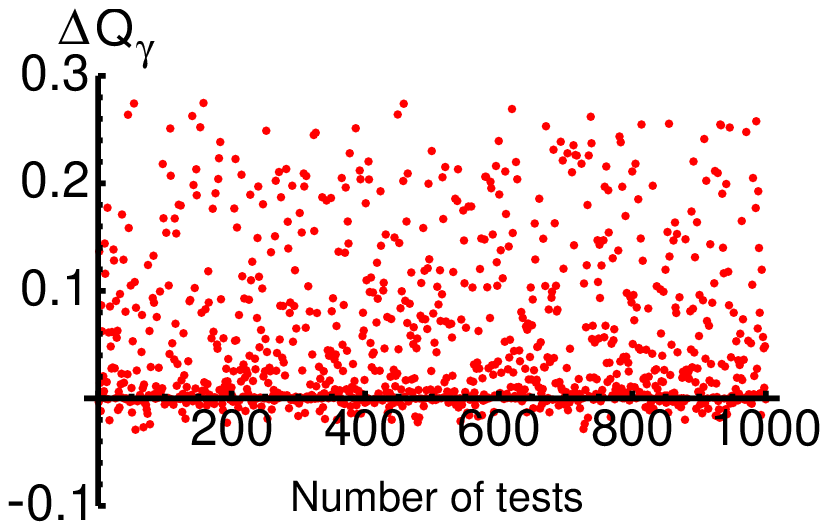} \caption{(Color online)
QCB reduction $\Delta Q_{\gamma}$ for different values of $\gamma$ (top left
$\gamma=0.99$, top right $\gamma=0.9$, bottom left $\gamma=0.8$, bottom right
$\gamma=0.7$). In each panel, $\Delta Q_{\gamma}$ is computed over a sample of
$10^{3}$ random values of $N$, $\beta$, and $\Gamma$. }%
\label{f:f5_gm}%
\end{figure}

Thus, we have just shown that, for fixed values of $N$ and $\beta$, the
asymmetric two-mode squeezed thermal state ($\gamma=1$) is the optimal thermal
probe in the detection of any loss $\Gamma$. Here we also show that this is
approximately true for $\gamma\lesssim1$. In other words, we show that the
inequality $Q_{2}(N,\beta,\gamma)<Q_{1}(N,\beta)$ is robust against
fluctuations of $\gamma$ below the optimal value $\gamma=1$. This property is
clearly important for practical implementations. To study this situation, let
us consider the $\gamma$-dependent QCB reduction
\begin{equation}
\Delta Q_{\gamma}=Q_{1}(N,\beta)-Q_{2}(N,\beta,\gamma)~. \label{g_dependent}%
\end{equation}
In Fig.~\ref{f:f5_gm} we have specified this quantity for different values of
the asymmetry parameter $\gamma$ (each panel refers to a different value of
$\gamma$). Then, for every chosen $\gamma$, we have computed $\Delta
Q_{\gamma}$ over a sample of $10^{3}$ random values of $N$, $\beta$, and
$\Gamma$ (in each panel). As one can see from the figure, the quantity $\Delta
Q_{\gamma}$ is approximately positive also when $\gamma$ is quite different
from the unity.
%%%%%%
%%%%%%

\section{Analysis of the correlations}

\label{qcbcorr} Since two-mode squeezed thermal states are able to outperform
the single-mode counterpart under several physical conditions, it is natural
to investigate this improvement directly in terms of the correlations of the
input state. The quantification of the correlations is realized by using the
entanglement, the quantum discord and the quantum mutual information. In order
to quantify the degree of entanglement of a two-mode Gaussian state, we can
use the logarithmic negativity. Let us consider a bipartite Gaussian state
with CM\ given in Eq.~(\ref{CM_2modes}). It is easy to derive the symplectic
eigenvalues of the partially transposed state. These are given by
\[
\tilde{d}_{\pm}=\sqrt{\frac{\tilde{\Delta}-\sqrt{\tilde{\Delta}^{2}-4I_{4}}%
}{2}}~,
\]
where $\tilde{\Delta}=I_{1}+I_{2}-2I_{3}$, and the symplectic invariants
$I_{1}$, $I_{2}$, $I_{3}$ and\ $I_{4}$ are defined in Eq.~(\ref{syINVARIANTS}%
). From the smallest of these symplectic eigenvalues, we can compute the
logarithmic negativity, which is equal to%
\[
E=\max\{0,-\log2\tilde{d}_{-}\}~.
\]
A bipartite Gaussian state is entangled iff $\tilde{d}_{-}<1/2$, so that the
logarithmic negativity gives positive values for all the entangled states and
$0$ otherwise.

The quantum discord is defined as the mismatch of two different quantum
analogues of classically equivalent expressions of the mutual information and
may be used to quantify quantum correlations in mixed separable states. For a
two-mode squeezed thermal state with CM as in Eq. (\ref{CM_2}), the quantum
discord may be written as \cite{GQDSCa}
\begin{align}
D=  &  h(\sqrt{I_{2}})-h(d_{-})-h(d_{+})\nonumber\\
&  +h\left(  \frac{\sqrt{I_{1}}+2\sqrt{I_{1}I_{2}}+2I_{3}}{1+2\sqrt{I_{2}}%
}\right)
\end{align}
where
\[
h(x)=\left(  x+\frac{1}{2}\right)  \log\left(  x+\frac{1}{2}\right)  -\left(
x-\frac{1}{2}\right)  \log\left(  x-\frac{1}{2}\right)
\]
is the binary Shannon entropy. We have that, for $0\leq D\leq1$, the state may
be either entangled or separable, whereas all the states with $D>1$ are
entangled \cite{GQDSCa,GQDSCb}.

Finally, the quantum mutual information, which quantifies the amount of total,
classical plus quantum, correlations, is given by $I=S(\rho_{A})+S(\rho
_{B})-S(\rho_{AB})$, where $S(\rho)=-\mathop{\text{Tr}}\nolimits[\rho\log
\rho]$ is the von Neumann entropy of the state $\rho$ and $\rho_{A(B)}%
=\mathop{\text{Tr}}\nolimits_{B(A)}[\rho_{AB}]$ are the partial traces over
the two subsystems. For a two-mode squeezed thermal state with CM as in
Eq.~(\ref{CM_2}) the quantum mutual information can be computed using the
formula
\[
I=\frac{1}{2}\left[  h(\sqrt{I_{1}})+h(\sqrt{I_{2}})-h(d_{+})-h(d_{-})\right]
\,.
\]
%%%%%%%%%%%%%%%%%%%%%%%%%%%%%%%%%%%
\begin{figure}[h]
\includegraphics[width=0.48\columnwidth]{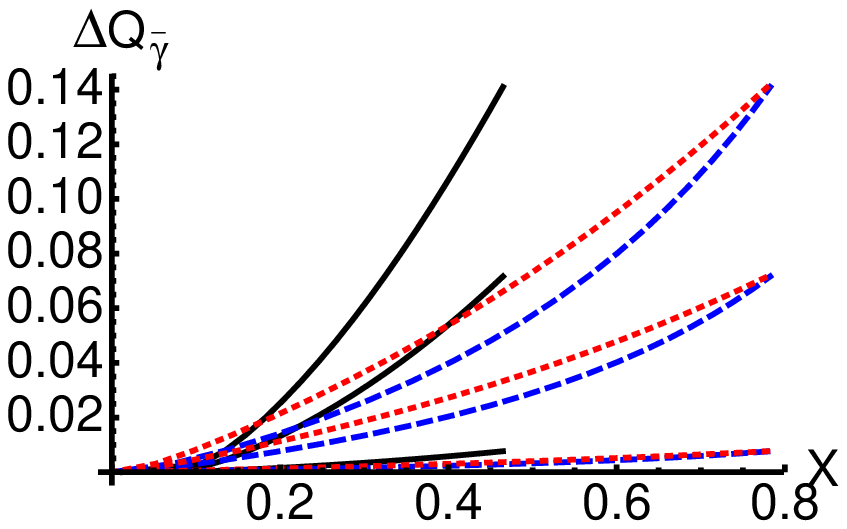}
\includegraphics[width=0.48\columnwidth]{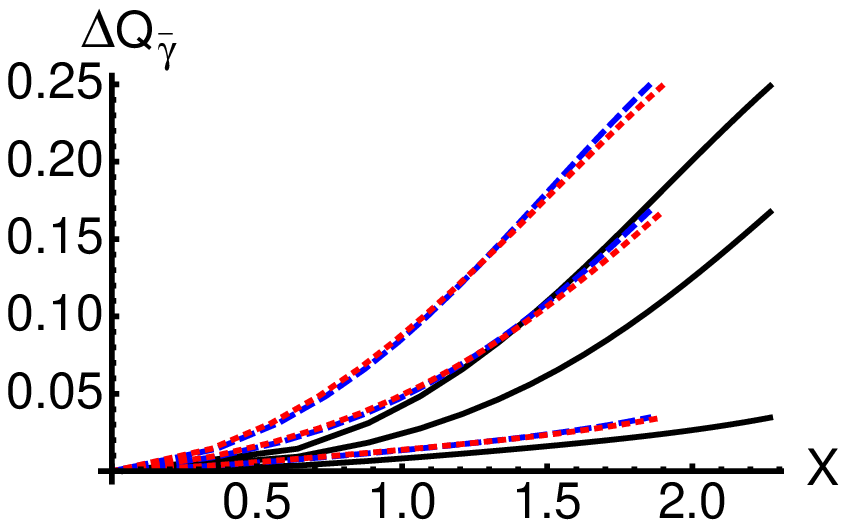}
\includegraphics[width=0.48\columnwidth]{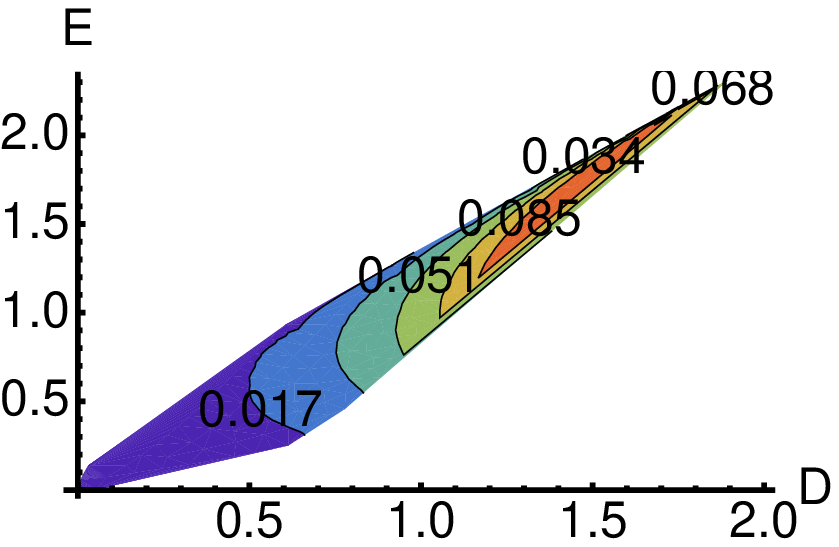}
\includegraphics[width=0.48\columnwidth]{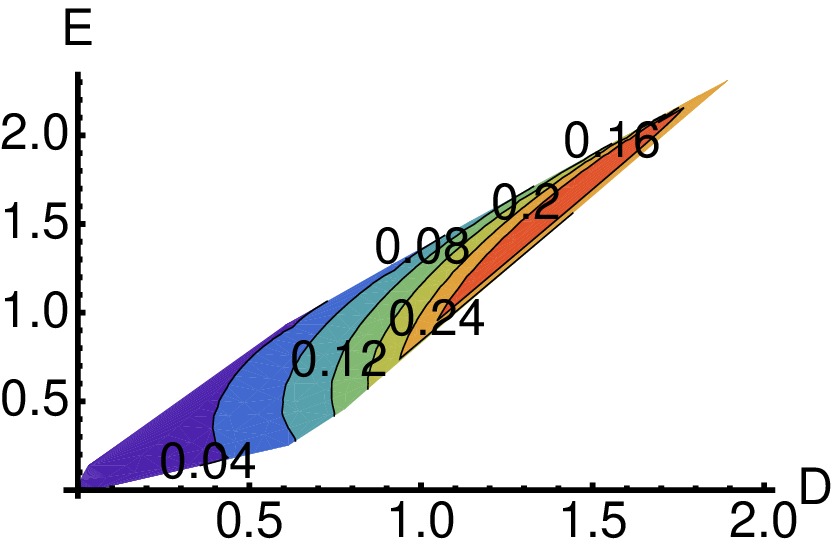}
\includegraphics[width=0.48\columnwidth]{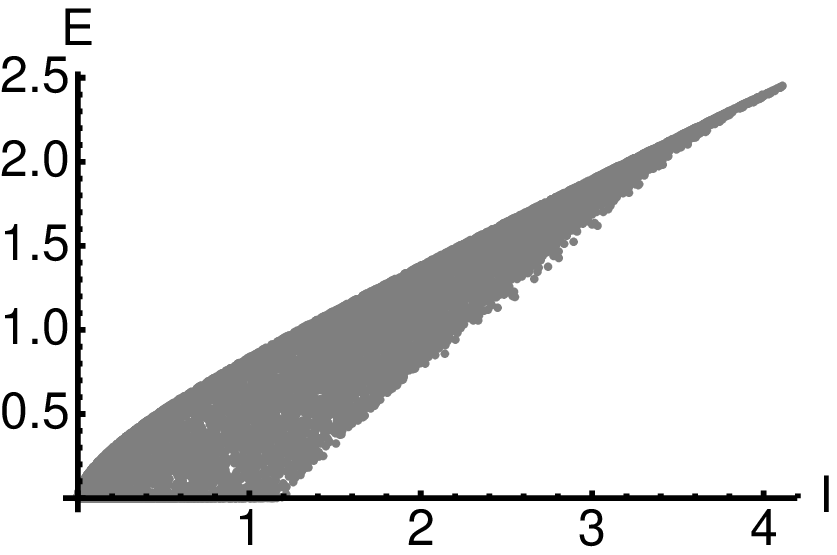}
\includegraphics[width=0.48\columnwidth]{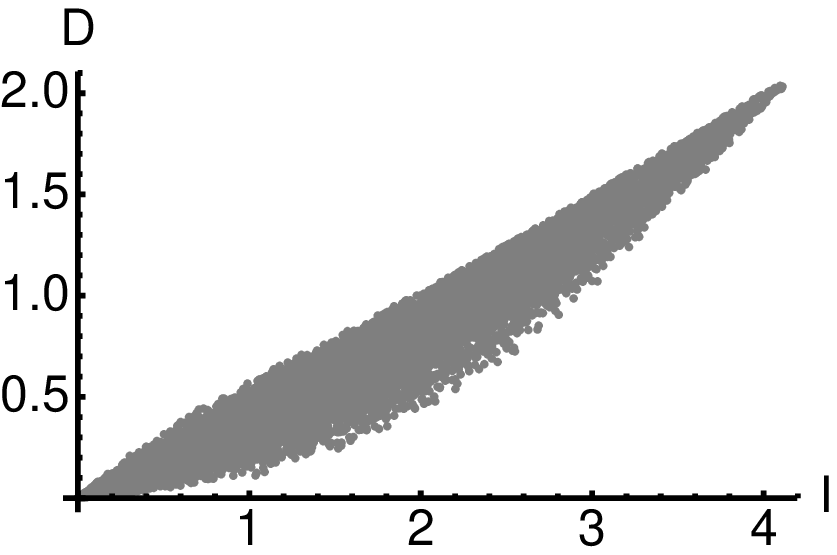}\caption{(Color online)
\textit{Upper panels}. QCB reduction $\Delta Q_{\bar{\gamma}}$ (with
$\bar{\gamma}=0.999$)\ as a function of the three correlation quantifiers
X=I,D,E where I is the quantum mutual information (dotted red), D is the
quantum discord (dashed blue) and E is the entanglement (solid black). The
plots are for fixed squeezing: $\beta=0.1$ for the left panel and $\beta=0.9$
for the right one. For each quantifier we plot three different curves
corresponding to different values of the damping (from top to bottom
$\Gamma=0.9,0.5$ and $0.1$). Each curve is generated by varying the input
energy $N$ between $0$ and $5$ photons.\textit{ Middle panels}. Density plots
of the QCB reduction $\Delta Q_{\bar{\gamma}}$ as a function of the input
discord and entanglement. The plots are for fixed damping: $\Gamma=0.2$ in the
left panel and $\Gamma=0.8$ in the right one. In each panel, the density plot
is generated by varying the squeezing $0\leq\beta\leq1$ and the energy $0\leq
N\leq5$. \textit{Lower panels}. Entanglement (left) and discord (right) as a
function of the quantum mutual information. Plots are generated by taking a
random sample of $10^{4}$ two-mode squeezed thermal states, i.e., random
values of $N$ and $\beta$ with $\gamma=\bar{\gamma}$.}%
\label{f:f6_cor}%
\end{figure}
%%%%%%%%%%%%%%%%%%%%%%%%%%%%%%%%%%%%%%%
For pure states the previous three measures are equivalent, whereas for mixed
states, as in the case under investigation in this section, they generally
quantify different kind of correlations. Here we consider the QCB\ reduction
$\Delta Q_{\bar{\gamma}}=Q_{1}(N,\beta)-Q_{2}(N,\beta,\bar{\gamma})$ between a
single-mode squeezed thermal state $\rho=\rho(N,\beta)$ and a two-mode
squeezed thermal state $\rho=\rho(N,\beta,\bar{\gamma})$ with $\bar{\gamma
}=0.999$. By fixing the input squeezing $\beta$ and varying the input energy
$N$, we study the behaviour of $\Delta Q_{\bar{\gamma}}$ as function of the
three correlation quantifiers, i.e., quantum mutual information, quantum
discord and entanglement (computed over the input two-mode state).

As shown in the upper panels of Fig.~\ref{f:f6_cor}, the QCB reduction $\Delta
Q_{\bar{\gamma}}$ is an increasing function of all the three correlation
quantifiers for fixed input squeezing ($\beta=0.1$ for the left panel and
$\beta=0.9$ for the right one). Note that, in each panel and for each
quantifier, we plot three different curves corresponding to different values
of the damping $\Gamma=0.9,$~$0.5$ and $0.1$. The monotonicity of the QCB
reduction in all the correlation quantifiers suggests that the presence of
correlations should definitely be considered as a resource for loss detection,
whether these correlations are classical or genuinely quantum, i.e., those
quantified by entanglement. In other words, employing the input squeezing in
the form of correlations is always beneficial for loss detection when we
consider squeezed thermal states as input sources. The importance of
correlations is confirmed by the plots in the middle panels. Here we consider
again the QCB reduction $\Delta Q_{\bar{\gamma}}=Q_{1}(N,\beta)-Q_{2}%
(N,\beta,\bar{\gamma})$ for $\bar{\gamma}=0.999$. Then, by varying input
squeezing $\beta$ and energy $N$, we study $\Delta Q_{\bar{\gamma}}$ as
function of both discord and entanglement (damping is $\Gamma=0.2$ in the left
panel, and $\Gamma=0.8$ in the right one). These plots show how the QCB
reduction is approximately an increasing function of both discord and
entanglement. Finally, in the lower panels of Fig.~\ref{f:f6_cor}, we also
show how entanglement (left) and discord (right) are increasing functions of
the quantum mutual information with good approximation (these plots are
generated by choosing a random sample of $10^{4}$ two-mode squeezed thermal states).

\section{Conclusions}

\label{conclusions} In this paper we have considered the quantum
discrimination of lossy channels. In particular, we have focused to the case
when one of the two channels is the identity, i.e., the problem of
discriminating the presence of a damping process from its absence (loss
detection). For this kind of discriminination we have considered thermal
probes as input, i.e., single- and two-mode squeezed thermal states. The
performance of the channel discrimination has been quantified using the QCB,
computed over the two possible states at the output of the unknown channel for
a given input state. Finding the optimal input state $\rho$ which minimizes
this bound gives automatically the optimal multi-copy state $\rho\otimes
\rho\otimes\cdots$ when we consider many accesses to the unknown channel
(under the assumption of single-copy constraints). In this scenario, we have
fixed the total energy of the input state and optimized the discrimination
(detection of loss) over the class of single- and two-mode squeezed thermal
states. We have found numerically that the optimal states are pure, thus
corresponding to single- and two-mode squeezed vacuum states. Furthermore, we
have determined the conditions where the two-mode state outperforms the
single-mode counterpart. This happens when the energy exceeds a certain
threshold, which becomes zero for suitably low values of the transmissivity
(i.e., high values of damping).

It is worth noticing that our approach (where we fix the total energy of
probing and reference modes) also gives a sufficient condition for the problem
where only the probing energy is fixed. In fact, if a two-mode state
outperforms a single-mode state above a certain threshold value $N_{th}$ of
the total energy, this also happens when just the energy of the probing mode
is above that value $N_{th}$. This is a trivial consequence of the fact that
the total energy is bigger than the probing energy for two-mode states
($N_{2}>N_{2}^{\mathrm{probe}}$) while the two quantities are the same for
single-mode states ($N_{1}=N_{1}^{\mathrm{probe}}$). Thus, $N_{2}%
^{\mathrm{probe}}=N_{1}^{\mathrm{probe}}>N_{th}$ can be written as
$N_{2}>N_{1}>N_{th}$ which is a stronger condition than $N_{2}=N_{1}>N_{th}$,
since the QCB is decreasing in the total energy, as one can see from
Eqs.~(\ref{QCB_1mode_ANA}) and~(\ref{QCB_2mode_ANA}).

In our investigation we have then considered the problem of loss detection in
more realistic conditions, where it is unlikely to have pure squeezing. In
this case, we have studied the optimal state for fixed total energy and
squeezing, i.e., by fixing all the relevant resources needed to create the
input state. Under these constraints, we have shown that a two-mode squeezed
thermal state which conveys all the thermal photons in the dissipative channel
is the optimal thermal probe. In addition, this result is robust against
fluctuations, i.e., it holds approximately also when the thermal photons are
distributed in a more balanced way between the probe mode (sent through the
dissipative channel) and the reference mode (bypassing the channel).

Finally we have closely investigated the role of correlations in our problem
of loss detection. We have found that, for fixed input squeezing, the
reduction of the QCB\ is an increasing function of several correlation
quantifiers, such as the quantum entanglement, the quantum discord and the
quantum mutual information. We then verify that employing the input squeezing
in the form of correlations (quantum or classical) is always beneficial for
the detection of loss by means of thermal probes.

The results of our paper provides new elements in the field of quantum channel
discrimination and can be applied to a wide range applications, including the
characterization of absorbing materials. In particular, they are relevant in
all the situations where the physical constraints regard the creation of the
input resources rather than the channel to be discriminated.

\section*{Acknowledgments}

The authors thank Marco Genoni, Stefano Olivares and Samuel L. Braunstein for
useful discussions.


\begin{thebibliography}{99}                                                                                               %


\bibitem {Breuer}H.-P. Breuer and F. Petruccione, \emph{The Theory of Open
Quantum Systems} (Oxford University Press, Oxford, 2002).

\bibitem {DecoRev}A. Serafini, M. G. A. Paris, F. Illuminati, and S. De Siena,
J. Opt. B \textbf{7}, R19 (2006).

\bibitem {Dau06}V. D'Auria, C. de Lisio, A. Porzio, S. Solimeno, and M. G. A.
Paris, J. Phys. B \textbf{39}, 1187 (2006).

\bibitem {Zurek}W. H. Zurek, Rev. Mod. Phys. \textbf{75}, 715 (2003).

\bibitem {Brask}J. B. Brask, I. Rigas, E. S. Polzik, U. L. Andersen, and A. S.
S\o {}rensen, Phys. Rev. Lett. \textbf{105}, 160501 (2010).

\bibitem {Jensen}K. Jensen, W. Wasilewski, H. Krauter, T. Fernholz, B. M.
Nielsen, M. Owari, M. B. Plenio, A. Serafini, M. M. Wolf and E. S. Polzik,
Nature Phys. \textbf{7}, 13 (2011).

\bibitem {Haroche08}M. Brune, J. Bernu, C. Guerlin, S. Del\'eglise, C. Sayrin,
S. Gleyzes, S. Kuhr, I. Dotsenko, J. M. Raimond, and S. Haroche, Phys. Rev.
Lett. \textbf{101}, 240402 (2008); S. Del\'eglise, I. Dotsenko, C. Sayrin, J.
Bernu, M. Brune, J. Raymond, and S. Haroche, Nature \textbf{455}, 510 (2008).

\bibitem {Wang}H. Wang, M. Hofheinz, M. Ansmann, R.C. Bialczak, E. Lucero, M.
Neeley, A. D. O'Connell, D. Sank, J. Wenner, A. N. Cleland, and John M.
Martinis, Phys. Rev. Lett. \textbf{101}, 240401 (2008).

\bibitem {Qtele}S. L. Braunstein, and H. J. Kimble, Phys. Rev. Lett.
\textbf{80}, 869 (1998); P. van Loock, and S. L. Braunstein, Phys. Rev. Lett.
\textbf{84}, 3482 (2000); S. Pirandola, S. Mancini, D. Vitali, and P. Tombesi,
Phys. Rev. A \textbf{68}, 062317 (2003); S. Pirandola, S. Mancini, D. Vitali,
and P. Tombesi, J. Mod. Opt. \textbf{51}, 901 (2004); S. Pirandola and S.
Mancini, Laser Physics \textbf{16}, 1418 (2006).

\bibitem {QKD1}C. Weedbrook, A. M. Lance, W. P. Bowen, T. Symul, T. C. Ralph,
and P. K. Lam, Phys. Rev. Lett.\textbf{ 93}, 170504 (2004); C. Weedbrook, S.
Pirandola, S. Lloyd, T. C. Ralph, Phys. Rev. Lett. \textbf{105}, 110501 (2010).

\bibitem {QKD2}S. Pirandola, S. L. Braunstein, and S. Lloyd, Phys. Rev. Lett.
\textbf{101}, 200504 (2008); S. Pirandola, S. Mancini, S. Lloyd, and S. L.
Braunstein, Nature Physics \textbf{4}, 726 (2008).

\bibitem {Devetak}I. Devetak and A. Winter, Phys. Rev. Lett. \textbf{93},
080501 (2004).

\bibitem {PirSKcapacity}S. Pirandola, R. Garc\'{\i}a-Patr\'{o}n, S. L.
Braunstein, and S. Lloyd, Phys. Rev. Lett. \textbf{102}, 050503 (2009); R.
Garc\'{\i}a-Patr\'{o}n, S. Pirandola, S. Lloyd, and J. H. Shapiro, Phys. Rev.
Lett. \textbf{102}, 210501 (2009).

\bibitem {Mon11}A. Monras and F. Illuminati, Phys. Rev. A \textbf{83}, 012315 (2011).

\bibitem {QreadingPRL}S. Pirandola, Phys. Rev. Lett. \textbf{106}, 090504 (2011).

\bibitem {NairLAST}R. Nair, \textquotedblleft Discriminating quantum optical
beam-splitter channels with number diagonal signal states: applications to
quantum reading and target detection\textquotedblright, arXiv:1105.4063.

\bibitem {Qillumination}S.-H. Tan, B. I. Erkmen, V. Giovannetti, S. Guha, S.
Lloyd, L. Maccone, S. Pirandola, and J. H. Shapiro, Phys. Rev. Lett.
\textbf{101}, 253601 (2008).

\bibitem {QilluminationOTHERS}S. Lloyd, Science \textbf{321}, 1463 (2008); J.
H. Shapiro and S. Lloyd, New J. Phys. \textbf{11}, 063045 (2009); H. P. Yuen
and R. Nair, Phys. Rev A \textbf{80}, 023816 (2009); S. Guha and B. Erkmen,
\textit{ibid.} \textbf{80}, 052310 (2009); A. R. Usha Devi and A. K.
Rajagopal, \textit{ibid.} \textbf{79}, 062320 (2009).

\bibitem {PRLtwo-mode}V. D'Auria, S. Fornaro, A. Porzio, S. Solimeno, S.
Olivares, and M. G. A. Paris, Phys. Rev. Lett. \textbf{102}, 020502 (2009).

\bibitem {Glauber}E. C. G. Sudarshan, Phys. Rev. Lett. \textbf{10}, 277
(1963); R. J. Glauber, Phys. Rev. \textbf{131}, 2766 (1963).

\bibitem {Freyberger}H. Venzl and M. Freyberger, Phys. Rev. A \textbf{75},
042322 (2007).

\bibitem {Mon07}A. Monras, M. G. A. Paris, Phys. Rev. Lett. \textbf{98},
160401 (2007).

\bibitem {AdessoR}G. Adesso, F. Dell'Anno, S. De Siena, F. Illuminati, and L.
A. M. Souza, Phys. Rev. A \textbf{79}, 040305(R) (2009).

\bibitem {Mon10}A. Monras and F. Illuminati, Phys. Rev. A \textbf{81}, 062326 (2010).

\bibitem {Helstrom}C. W. Helstrom, \emph{Quantum Detection and Estimation
Theory} (Academic Press, New York, 1976).

\bibitem {Chefles}A. Chefles, Contemp. Phys. \textbf{41}, 401 (2000).

\bibitem {QSE}J. A. Bergou, U. Herzog, M. Hillery in \emph{Quantum State
Estimation}, Lect. Not. Phys. \textbf{649}, J. Rehacek, M. G. A. Paris (Eds)
(Springer, Berlin, 2004), pp 417-465.

\bibitem {QSE2}A. Chefles in \emph{Quantum State Estimation}, Lect. Not. Phys.
\textbf{649}, J. Rehacek, M. G. A. Paris (Eds) (Springer, Berlin, 2004), pp 467-511.

\bibitem {Kargin}V. Kargin, Ann. Stat. \textbf{33}, 959 (2005).

\bibitem {PRAQCB}J. Calsamiglia, R. Munoz-Tapia, L. Masanes, A. Ac\'{\i}n, E.
Bagan, Phys. Rev. A \textbf{77}, 032311 (2008).

\bibitem {PRLQCB}K. M. R. Audenaert, J. Calsamiglia, R. Munoz-Tapia, E. Bagan,
Ll. Masanes, A. Acin, F. Verstraete, Phys. Rev. Lett. \textbf{98}, 160501 (2007).

\bibitem {Nuss}M. Nussbaum, and A. Szkola, Ann. Stat. \textbf{37}, 1040 (2009).

\bibitem {Aud}K. M. R. Audenaert, M. Nussbaum, A. Szkola, and F. Verstraete,
Commun. Math. Phys. \textbf{279}, 251 (2008).

\bibitem {Pirandola}S. Pirandola and S. Lloyd, Phys. Rev. A \textbf{78},
012331 (2008).

\bibitem {Disc1}H. Ollivier, W. H. Zurek, Phys. Rev. Lett. \textbf{88}, 017901
(2001); W. H. Zurek, Phys. Rev. A \textbf{67}, 012320 (2003).

\bibitem {Disc2}C. A. Rodriguez-Rosario, K. Modi, A. Kuah, A. Shaji, and E. C.
G. Sudarshan, J. Phys. A \textbf{41}, 205301 (2008); M. Piani, P. Horodecki,
and R. Horodecki, Phis. Rev. Lett. \textbf{100}, 090502 (2008).

\bibitem {Ferraro10}A. Ferraro, L. Aolita, D. Cavalcanti, F. M. Cucchietti,
and A. Acin, Phys. Rev. A \textbf{81}, 052318 (2010).

\bibitem {GQDSCa}P. Giorda and M. G. A. Paris, Phys. Rev. Lett. \textbf{105},
020503 (2010).

\bibitem {GQDSCb}G. Adesso and A. Datta, Phys. Rev. Lett. \textbf{105}, 030501 (2010).

\bibitem {Vedral}V. Vedral, Rev. Mod. Phys. \textbf{74}, 197 (2002).

\bibitem {Boca}M. Boca, I. Ghiu, P. Marian, and T. A. Marian, Phys. Rev. A
\textbf{79}, 014302 (2009).

\bibitem {Ghiu}I. Ghiu, G. Bj\"ork, P. Marian, and T. A. Marian, Phys. Rev. A
\textbf{82}, 023803 (2010).

\bibitem {AbastoQCM}D. F. Abasto, N. T. Jacobson, P. Zanardi, Phys. Rev. A
\textbf{77}, 022327 (2008).

\bibitem {JMOQCB}C. Invernizzi, M. G. A. Paris, J. Mod. Opt \textbf{57}, 1362 (2010).

\bibitem {GSinQI}A. Ferraro, S. Olivares, M. G. A. Paris, \emph{Gaussian
States in Quantum Information} (Bibliopolis, Napoli 2005).

\bibitem {QITCV}S. L. Braunstein and A. K. Pati, \emph{Quantum Information
Theory with Continuous Variables} (Kluwer Academic, Dordrecht, 2003).

\bibitem {Fuchs}C.A. Fuchs and J. V. de Graafs, IEEE Trans. Inf. Theory
\textbf{45}, 1216 (1999).

\bibitem {Nielsen}M. A. Nielsen and I. L. Chuang,\textit{ Quantum Computation
and Quantum Information} (Cambridge University Press, Cambridge, 2000).

\bibitem {QreadingSUPP}S. Pirandola, Phys. Rev. Lett. \textbf{106}, 090504
(2011): Online Supplementary Material (http://link.aps.org/ supplemental/10.1103/PhysRevLett.106.090504).

\bibitem {Segno}Note that, in our derivations, what is important is the
modulus of the squeezing parameter but not its sign. As a result, we can
choose $\zeta=-r$ as well as $\zeta=r$.

\bibitem {DiagSympl}A. Serafini, F. Illuminati, and S. de Siena, J. Phys. B
\textbf{37}, L21 (2004).

\bibitem {PirCMs}S. Pirandola, A. Serafini, and S. Lloyd, Phys. Rev. A
\textbf{79}, 052327 (2009).

\bibitem {ContQCB}Note that, because of the continuity of the QCB, the error
probability in the discrimination between a lossy channel ($\Gamma>0$) and an
almost-ideal channel ($\Gamma=\varepsilon\simeq0$) can be made arbitrarily
close to the error probability affecting the discrimination of the same lossy
channel ($\Gamma>0$) and an ideal channel ($\Gamma=0$).

\bibitem {Walls}D. F. Walls and G. J. Milburn, \emph{Quantum Optics}
(Springer, Berlin, 1994).

\bibitem {AS1}A. Serafini, F. Illuminati, M. G. A. Paris, S. De Siena, Phys.
Rev A \textbf{69}, 022318 (2004).

\bibitem {OtherCurve}Clearly we can invert the curve and introduce a threshold
transmissivity as function of the energy $\eta_{th}=\eta_{th}(N)$.
For values $\eta<\eta_{th}$\ the two-mode state is better than the
single-mode state, while the opposite happens for
$\eta>\eta_{th}$. We have $\eta_{th}\simeq \eta_{c}+0.18\,N^{0.7}$
for small $N$ and $\eta_{th}\simeq1-2/N$ for large $N$.


\end{thebibliography}
\end{document}